\documentclass{lmcs}
\pdfoutput=1

\usepackage{lastpage}
\lmcsdoi{17}{4}{11}
\lmcsheading{}{\pageref{LastPage}}{}{}%
{Nov.~02,~2017}{Nov.~24,~2021}{}

\usepackage[utf8]{inputenc}

\bibstyle{alpha}
\usepackage{oldlfont,amssymb,latexsym,amsfonts,euscript,stmaryrd}
\input{xy}
\xyoption{all}

\def\[{\ensuremath{[ \! [}}
\def\]{\ensuremath{] \! ]}}
\def\({\ensuremath{( \! [}}
\def\){\ensuremath{] \! )}}
\def\succ{{\mathsf{suc}}}
\def\merge{{\mathtt{merge}}}
\newcommand{\unmerge}{{\mathtt{unmerge}}}

\def\ocd{{\mathcal{OC}}}
\newcommand{\ocde}{{\mathcal{OC_E}}}
\newcommand{\ocds}{{\mathcal{OC_S}}}
\newcommand{\U}{{\mathcal{U}}}
\newcommand{\fun}{{\mathsf{fun}}}
 \newcommand{\bchoice}{{\mathtt{?}}}
\newcommand{\eq}{{\mathtt{Eq}}}

\def\o{{\mathit{o}}}
\newcommand{\coin}{{\mathtt{coin}}}
\newcommand{\toss}{{\mathtt{toss}}}
\newcommand{\newcoin}{{\mathtt{new\_coin}}}

\newcommand{\e}{\mho}
\newcommand{\fst}{{\mathtt{fst}}}
\newcommand{\snd}{{\mathtt{snd}}}
\newcommand{\pair}{{\mathtt{pair}}}

\newcommand{\sbot}{\bot_S}
\newcommand{\ebot}{\bot_E}
\newcommand{\tops}{\top_S}
\newcommand{\etop}{\top_E}
\newcommand{\espcf}{$\Lambda^+(N)\spc$}
\newcommand{\nt}{{\mathit{N}}}
\newcommand{\xp}{{\cal{P}}}
\newcommand{\strat}{{\mathsf{strat}}}

\def\Y{{\mathbf{Y}}}

\def\0{{\mathbf{0}}}

\def\1m{\iota}

\def\1{{{\mathbf{1}}}}

\def\B{{\mathcal{B}}}
\newcommand{\EB}{{\mathcal{B_E}}}
\newcommand{\SB}{{\mathcal{B_S}}}
\newcommand{\si}{{\mathsf{si}}}

\newcommand{\sqleq}{\sqsubseteq_E}

\newcommand{\leqs}{\leq_S}

\newcommand{\vvdash}{\vdash^*}

\newcommand{\may}{{\mathit{may}}}
\newcommand{\must}{{\mathit{must}}}
\newcommand{\lmay}{{\lesssim^{\may}}}
\newcommand{\lmust}{{\lesssim^{\must}}}

\newcommand{\rnd}{{?_\nat}}

\newcommand{\proj}{{\mathtt proj}}

\newcommand{\pj}{{\mathtt{proj}}}
\newcommand{\orr}{+}

\newcommand{\spc}{\hspace{2pt}}

\newcommand{\up}{\uparrow}

\newcommand{\callcc}{{\mathtt{callcc}}}

\newcommand{\nat}{{\mathtt{nat}}}

\newcommand{\If}{{\mathtt{If}}}

\newcommand{\hd}{{\mathtt{hd}}}
\newcommand{\tl}{{\mathtt{tl}}}

\newcommand{\Na}{{\mathbb{N}}}

\newcommand{\boo}{{\mathtt{bool}}}

\newcommand{\ff}{{\mathtt{ff}}}
\newcommand{\tru}{{\mathtt{tt}}}
\newcommand{\pred}{{\mathtt{pred}}}
\newcommand{\then}{{ {\mathtt{then}} }}
\newcommand{\el}{{{\mathtt{else} }}}

\newcommand{\suc}{{\mathtt{suc}}}
\newcommand{\K}{{\cal{K}}}

\newcommand{\inj}{{\mathtt{inj}}}

\bibstyle{alpha}
\begin{document}
%\begin{frontmatter}
  \title[Extensional and Intensional Models for Nondeterminism]{Extensional and Intensional Models for Bounded and Unbounded Nondeterminism}
  \author[J.~Laird]{J. Laird}
\address{Department of Computer Science, University of Bath, UK}
%\maketitle
 \def\o{{\mathit{o}}}
\begin{abstract} 
We give extensional and intensional characterizations of functional programs with nondeterminism: as structure preserving functions between  \emph{biorders}, and as nondeterministic \emph{sequential algorithms} on  \emph{ordered concrete data structures} which compute them.  
%By combining the extensional (Scott) and stable (Berry) orders, biorders allow an extensional characterization of sequential functions, and also a satisfactory interpretation of unbounded non-determinism.    
A fundamental result establishes that these extensional and intensional representations  are equivalent, by showing how to construct the unique sequential algorithm  which computes a given monotone and stable function, and describing the conditions on sequential algorithms which correspond to continuity with respect to each order.

We illustrate  by defining may-testing and must-testing denotational semantics for sequential functional languages with bounded and unbounded choice operators. We prove that these are computationally adequate, despite the non-continuity of the must-testing semantics of unbounded nondeterminism. % in which recursive functions are interpreted as least fixed points of transfinite chains of approximants: here the stable order proves its value in giving a satisfactory account of unbounded non-determinism. 

In the bounded case, we prove that our continuous  models are fully abstract with respect to may-testing and must-testing by identifying a simple universal type, which may also form the basis for models of the untyped $\lambda$-calculus. In the unbounded case we observe that our model contains computable functions which are not denoted by terms, by identifying a further ``weak continuity'' property  of the definable elements, and use this to establish that it is not fully abstract. 

\end{abstract}
\keywords{Nondeterminism, Denotational Semantics, Biorders, Sequential algorithms}
\subjclass{F.3.2[Semantics of Programming Languages]}
\titlecomment{}
\maketitle
%\end{keyword}
%\end{frontmatter}
\section{Introduction}
\label{sec:s1}
This paper describes denotational models of sequential higher-order functional computation with nondeterminism,  which may be given as an  explicit choice operation or arise via an under-specified operational semantics or  abstract interpretation of a richer language. Interpreting  nondeterminism in a sequential setting presents some novel challenges for denotational semantics --- for example, accurately representing the branching points within a program at which choices are made. Moreover \emph{unbounded} nondeterminism  allows programs to be written which will always return a value but may take an unbounded number of steps to do so (corresponding to the notion of \emph{fairness} \cite{djikstra}). Any sound denotational interpretation of such programs will not be a continuous function, meaning that many standard semantic techniques are not available.

We take an approach which relates  extensional and {intensional} representations of nondeterministic functional programs, by developing the theory of \emph{biorders} (sets with two partial orders) and applying it to nondeterminism, and  defining a new notion of \emph{ordered concrete data structure} --- a game-like structure representing the interaction between a program and its environment at a given type. 
Our key results are based on an isomorphism between the \emph{monotone stable}  functions between the biorders of states of  ordered concrete data structures, and the sequential algorithms   which compute them. We show that these equivalent representations  may be used to give may-and-must testing models of an ``observably sequential'' functional programming language  with nondeterminism,  \espcf. These models are fully abstract when restricted to bounded choice, as is shown by identifying a simple universal type, of which every other computational type is a retract. However, the  model of must testing for unbounded non-determinism  contains a class of elements which are not definable as terms, leading to a failure of full abstraction and suggesting that it could be further refined by restricting to ``weakly continuous'' functions.

\subsection{Related Work:}
 
{\bf{Extensional Semantics for Non-determinism.}} There is a body of research describing  the appropriate order-theoretic structures for representing nondeterminism denotationally, and establishing their relationship to notions of testing and equivalence~--- e.g. \cite{HA,Plotp}. The principal difficulty which arises when interpreting countable nondeterminism~--- the non-continuous nature of such a semantics~---  may be resolved by weakening the continuity properties demanded (e.g. to $\omega_1$-continuity \cite{AP}). However, this  admits  many undefinable functions  and leaves fewer principles with which to reason about program behaviour (not least, proving computational adequacy). 

In this work we focus on the relationship between sequentiality of (higher-order) functions and non-determinism, axiomatizing the structure required to interpret choice in this setting.  More specifically,  our approach is based on two orders~--- an extensional order corresponding to the observable  input/output behaviour of functions, and a  \emph{stable} order, capturing, in effect, the minimal computation required to produce a given result. This is  foreshadowed by Roscoe's work on the semantics of CSP (in which the stable order is called the ``strong order'') \cite{Roscoe1,Roscoe2} and Berry's introduction of \emph{biorders} \cite{Be}, which combine both orders in constructing models of functional (but deterministic) languages.  
  In  previous work, the author has observed that stable and continuous functions on biorders with a (extensionally) greatest element are  sequential in the sense of Milner and Vuillemin, and used them to construct fully abstract  models of sequential languages such as the lazy $\lambda$-calculus \cite{ufpc}. However, although  these models technically carry information about the way  higher-order  programs evaluate their arguments, this must be recovered from the graphs or traces of the functions they denote, suggesting that an alternative representation would be valuable for program analysis. 

One  approach   is suggested by the  work of  Curien,  Plotkin and  Winskel \cite{cpw}, which describes a fully faithful functor into the category of bidomains from  a Cartesian closed category constructed from a model of linear logic based on \emph{bistructures}. This is a more general representation of bidomains than that given by ordered concrete data structures (i.e. the biorders which are finite-branching states of the latter are strictly included in those which correspond to cliques of the former).  Thus it includes stable and extensionally continuous functions which are not sequential, such as Gustav's function \cite{Be}. As a continuous semantics, it does not capture programs with unbounded nondeterminism.

\paragraph*{Intensional Semantics for  Non-determinism.} Given the success of intensional techniques~--- in particular, game semantics~---  in characterizing deterministic sequential computation through fully abstract models of programming languages (with and without side-effects), extending these models with non-determinism by relaxing the determinacy constraint on strategies is a natural step. This was  taken by Harmer and McCusker \cite{HM}, who gave a game semantics of Idealized Algol with  bounded non-determinism, fully abstract with respect to may-and-must testing equivalence. This bears clear parallels with our intensional  model, e.g. in the explicit recording of divergent traces.  
Representing the branching behaviour of functional programs without state is more challenging; witness the difficulty of  combining innocence with nondeterminism \cite{hart}, resolved in work by Tsukada and Ong \cite{TO} based on a sheaved model, as well as in concurrent games models \cite{RW} This succeeds in  characterizing the definable elements in a model of the $\lambda$-calculus with choice, although it does not capture their observational equivalence in a direct or effective manner.  Intensional models of  unbounded nondeterminism encounter challenges related to non-continuity~--- as noted in \cite{HM}, representations of strategies as collections of finite sequences of moves are insufficient to capture the distinction between infinite interactions and finite, but unbounded ones, and Levy \cite{inftrace} describes a semantics of recursively defined nondeterministic programs in terms of their infinite traces. Our model similarly contains representations of infinitary interaction, although these are positions which may be reached by (potentially, many different) ordinal chains of events.

The  intensional semantics for nondeterminism developed here is based on \emph{concrete data structures}, which were introduced by Kahn and Plotkin \cite{KP}, as part of a definition of sequentiality for (deterministic) higher-order functionals (see further discussion in Section~6.2), although they offer an appealing model of computation in their own right, via the notion of \emph{sequential algorithm} \cite{BeCu}.
 On the one hand, concrete data structures correspond to a  \emph{positional} form of games, and sequential algorithms to positional strategies  (see e.g. \cite{HyS}). On the other, sequential algorithms  may be related to purely extensional models: in the deterministic case, Cartwright, Curien and Felleisen \cite{CCF} have established  that they compute, and are equivalent to ``observably sequential'' functions; the author has given a more abstract representation of the latter as  \emph{bistable} functions on bistable biorders  \cite{bist,lbd}.

Berry and Curien's deterministic sequential algorithms are \emph{strongly sequential}~--- in each position the  data required to compute further is \emph{unique} (each \emph{cell} is filled with a unique \emph{value}). By contrast, we describe a  \emph{weakly sequential} representation (cells may be filled with multiple values)~---  which may be required either because evaluation is explicitly non-deterministic, or because evaluation order is not observable~---   by abandoning this consistency condition.  However, this also  requires an ordering on cells and values (corresponding to game positions), to reflect the fact that (for example) any program which may diverge in response to a given argument may still diverge in response to an argument with a wider range of behaviours.  This notion of an ordered concrete data structure was first introduced in \cite{csl09}, in which stable and continuous functions were shown to correspond to \emph{finite-branching} sequential algorithms. Extension to unbounded nondeterminism \cite{mfps15}  requires a new notion of transfinite (ordinal-indexed) interaction, to distinguish computations which are infinite from those which are finite but unbounded.

\subsection{Outline of the Paper}
Section~\ref{sec:s2} recalls and brings together some earlier work by using it to give a biorder  semantics of a minimal functional language with non-determinism, $\Lambda^+$. This illustrates some features of sequential nondeterministic higher-order functions, and motivates the rest of the paper which expands and elaborates on this model. Section~\ref{sec:s3} extends it  with infinitary data (natural numbers)  and fixed points, allowing the expression of countable nondeterminism, and shows that these may be interpreted soundly by introducing notions of completeness to biorders. 
Section~\ref{sec:s4} introduces \emph{ordered concrete data structures} (ocds) and  their biorders of states, and  Section~\ref{sec:s5} introduces non-deterministic sequential algorithms and shows that they compute monotone stable functions between biorders. Section~\ref{sec:s6} extends these results to show that each such function is computed by a unique sequential algorithm, and thus establish Cartesian closure for our categories of ocds.
Section~\ref{sec:s7} returns to the semantics of nondeterminism, proving full abstraction and universality results for the models of bounded nondeterminism, and describing a further ``weak  continuity'' property of definable elements in the must testing semantics of unbounded nondeterminism, leading to a proof that it is not fully abstract. 

\section{Non-determinism and Weak Sequentiality}
\label{sec:s2}
Let $\Lambda^+$ be the simply-typed $\lambda$-calculus over a single base type $\o$ with two ground type constants~--- $\Omega:\o$ (representing divergence)  and $\mho:\o$ (representing immediate termination, due to error, deadlock etc.)~--- and a binary choice operation $\orr: \o \rightarrow \o \rightarrow \o$, which we write infix. Programs (closed terms of type $\o$) may be evaluated by (leftmost outermost) reduction) of $\lambda x.r\spc s$ to $r[s/x]$ and  $r \orr s$  to either $r$ or $s$. If every such reduction of a program $t:\o$ terminates with $\mho$ then we say that $t$ \emph{must} converge ($t \Downarrow^\must$) and if some reduction of $t$  terminates with $\mho$ then we say that $t$ \emph{may} converge ($t \Downarrow^\may$).

Alternatively, these may and must convergence predicates may be defined  via the  ``big step'' reduction rules in Table \ref{opsem1} (in which the rules in the top row apply to both may and must convergence predicates). 
\begin{table}
\begin{center}
\begin{tabular}{c c c}
{\Large $\over \mho \Downarrow \mho$} {\Large $\over \lambda x.t \Downarrow \lambda x.s$} & & {\Large $s \Downarrow \lambda x.s'\ \ \  s'[t/x] \Downarrow C \over  s\spc t \Downarrow C$}\\\\ 
{\Large $s \Downarrow^\must \mho \ \ \ \ t \Downarrow^\must \mho  \over s \orr t \Downarrow^\must \mho$} & & {\Large $s \Downarrow^\may \mho \over s \orr t \Downarrow^\may\mho $}\  {\Large $t \Downarrow^\may \mho \over s \orr t \Downarrow^\may\mho $} \\\\
\end{tabular}
\caption{Big-step rules for May and Must Testing}\label{opsem1}
\end{center}
\end{table}
%where $C$ is either a $\lambda$-abstraction, or $\mho$. 
 They are used as the basis for the following notions of observational approximation on terms of $\Lambda^+$:
\begin{description}
\item [May  Testing] $s \lesssim^\may  t$ if for all compatible program contexts $C[\cdot]:\o$, $C[s]\Downarrow^\may \mho$ implies $C[t]\Downarrow^\may \mho$. %  $M \simeq^\may N$ if $M \lesssim^\may N$ and  $N \lesssim^\may M$.
 \item [Must Testing]$s \lesssim_\must t$ if for all compatible program contexts $C[\cdot]:\o$, $C[s] \Downarrow_\must \mho $ implies $C[t]\Downarrow_\must \mho$. % $M \simeq_\must N$ if $M \lesssim_\must N$ and  $N \lesssim_\must M$.
 \item [May-and-Must Testing]$s \lesssim^\may_\must t$ if $s \lesssim^\may  t$  and  $s \lesssim_\must t$. % $M \simeq_\must^\may N$ if $M \lesssim_\must^\may N$ and  $N \lesssim_\must^\may M$.
\end{description}
We obtain an equivalence by the intersection of each approximation relation with its converse.  
%Since may-and-must equivalence is simply the conjunction of may-equivalence and must equivalence 
%Note that these notions of 
For example, let  $\boo$ be the type $\o \rightarrow \o \rightarrow \o$, which contains as its equivalence classes (in each case) the values $\tru \triangleq \lambda x.\lambda y.x$ and $\ff \triangleq \lambda x.\lambda y.y$,  the error and divergence terms $\mho_\boo \triangleq \lambda x.\lambda y.\mho$ and $\Omega_\boo \triangleq \lambda x.\lambda y.\Omega$  and the non-deterministic choice $\tru + \ff \triangleq \lambda x.\lambda y.(x +y)$. We have $\Omega \lesssim^\may \tru,\ff \lesssim^\may \tru \orr \ff \lesssim^\may \mho$ and  $\Omega \lesssim_\must \tru \orr \ff \lesssim_\must  \tru,\ff  \lesssim_\must \mho$.

We make the following observations:
\begin{itemize}
\item The notions of approximation and equivalence are  extensional, in the sense that closed programs of function type will be observationally equivalent if and only if they are equivalent when applied to any closed term. So, in particular, there are only finitely many equivalence classes at each type.
\item In this finitary setting, may testing and must testing (and thus the corresponding notions of approximation and equivalence) are dual, in the sense that $t \Downarrow^\may$ if and only if  $t[\Omega/\mho, \mho/\Omega]\Downarrow^\must$. So one model will suffice for both.
\item The operational semantics for must testing is  the same as that given  for  \emph{Unary PCF} in e.g. \cite{Lou}. Unary PCF is the simply-typed $\lambda$-calculus over a ground type containing a single value, and a single binary operator on this type which tests both of its arguments for convergence to that value. The latter is usually regarded as a sequential composition, but since the order of evaluation is not observable, the arguments may be evaluated in either order, or in parallel, or non-deterministically. 
 \end{itemize}
By relating non-determinism to Unary PCF we may use some  important facts about the latter --- including that its observational equivalence is decidable \cite{Lou}, and that it has a universal model in the category of biorders and monotone stable functions \cite{ufpc}.  Conversely, our sequential algorithms model may be seen as a solution to the problem of giving a corresponding intensional semantics of Unary PCF.

Unary PCF (and thus $\Lambda^+$)  is a sequential language in the sense of Milner-Vuillemin: for any term $t:T_1 \rightarrow \ldots T_k \rightarrow \o$ which  is strict (i.e. there is some sequence of arguments on which it diverges) there is a \emph{sequentiality index} $i \in \{1,\ldots,k\}$ such that  $t\spc s_1\ldots s_k \Downarrow^\must$ implies $s_i \not \simeq_\must  \lambda \vec{x}.\Omega$. In fact, it is \emph{weakly} sequential --- there are strict but non-divergent terms with more than one sequentiality index --- e.g. $\lambda xy.x+y$.   
As in this instance, weak sequentiality may arise because evaluation is non-deterministic, or because the order of evaluation is underspecified or unobservable.  
Other examples of weakly sequential languages include PCF itself (for which the fully abstract model is known to not be effectively presentable \cite{Lo}), and the  call-by-value and lazy $\lambda$-calculi \cite{apb} --- e.g. in the absence of side-effects such as mutable state, a call-by-value application $t\spc s$ may be evaluated by first evaluating $t$ to a value, then $s$, or vice-versa (or, indeed, non-deterministically or  in parallel)~--- which do have effectively presentable fully abstract models based on biorders \cite{ufpc}.

 Intensional (game) semantics are typically based on explicitly sequential representations of interaction such as sequences of moves: one of the aims of the current work is to investigate a model of higher-order computation which is intrinsically weakly sequential. The following example illustrates  why this is needed to give  fully abstract semantics of  non-deterministic functional programs.
\begin{exa}\label{nfa}
 Define the following terms $x_0:\boo,x_1:\boo \vdash  t_{ij}: \boo$
\begin{itemize}
\item $t_{0,1} \triangleq  \If\spc x_0\spc \then \spc x_1\spc \el\spc x_1$, (i.e. $\lambda ab.(x_0\spc ((x_1\spc a)\spc b)) \spc ((x_1 \spc a)\spc b)$)
\item $t_{1,0} \triangleq \If\spc x_1\spc \then \spc x_0\spc \el\spc x_0$, %\lambda xyab.(y\spc ((x\spc a)\spc b)) \spc ((x \spc a)\spc b)$ (i.e. $ 
\item $t_{0,0} \triangleq \If\spc x_0\spc \then \spc (\If\spc x_1\spc \then\spc \tru\spc \el \spc \tru)\spc \el\spc (\If\spc x_1\spc \then\spc \ff\spc \el \spc \ff)$, %= \lambda xyab.(x\spc ((y\spc a)\spc a)) \spc ((y \spc b)\spc b)$ (i.e. $
\item $t_{1,1} \triangleq \If\spc yx_1\spc \then \spc (\If\spc x_0\spc \then\spc \tru\spc \el \spc \tru)\spc \el\spc (\If\spc x_0\spc \then\spc \ff\spc \el \spc \ff)$.%\lambda xyab.(y\spc ((x\spc a)\spc a)) \spc ((x \spc b)\spc b)$ (i.e. $\
\end{itemize}
Each term evaluates both arguments and returns one of them --- $t_{ij}$ evaluates argument $i$ first and returns the value of argument $j$. Thus they are pairwise  (may or must) observationally distinguishable --- $t_{0,1}$ and $t_{0,0}$ may be distinguished from  $t_{1,0}$ and $t_{1,1}$  by application to $\Omega$ and $\mho$, and  $t_{1,0}$ and $t_{0,0}$  may be distinguished from $t_{0,1}$ and $t_{1,1}$  by application to $\tru$ and $\ff$.
  However the terms $t_{0,1} \orr t_{1,0}$  and $t_{0,0} \orr t_{1,1}$ are may-and-must equivalent --- both can test their  arguments in either order,  and  return either one of them. Note that they are  observationally distinguishable using mutable state (e.g. in erratic Idealized Algol), by using an imperative variable to record the order in which the arguments \emph{actually were} tested. 
\end{exa}
%This example suggests that a strongly sequential representation of program-argument interaction  is not what is needed to give a fully abstract model of $\Lambda^+$.

A second (well-known) example illustrates another difficulty of giving an intensional semantics of non-deterministic functions --- in addition to capturing the order in which they explore their arguments, it is necessary to represent the \emph{branching points} where different evaluation paths may be chosen.  
\begin{exa}\label{bta}
The following approximations are strict:
\begin{center} $\lambda f.(f\spc \tru) \orr (f\spc \ff)  \lesssim^\may \lambda f.f\spc (\tru \orr \ff)$ and 
 $\lambda f.f\spc (\tru \orr \ff)  \lesssim_{\must} \lambda f.(f\spc \tru) \orr (f\spc \ff)$  \end{center}
Indeed,  these terms are  separated by application to $\lambda x. \If\spc x\spc \then\spc \Omega \spc \el \spc (\If \spc x\spc \then\spc \mho \spc \el \spc \Omega)$ (may-testing) and  $\lambda x. \If\spc x\spc \then\spc \mho \spc \el \spc (\If \spc x\spc \then\spc \Omega \spc \el \spc \mho)$ (must-testing).

 To capture the  difference between  these functions at the level of the strategies which compute them, it is necessary to record whether different calls by $f$ to its argument may be  supplied with different values. For strategies which are innocent (i.e. not stateful) this requires  a representation of  interaction which captures branching points, such as the sheaved model of \cite{TO}
\end{exa}

\subsection{Biorders and Stable Functions}
Turning to denotational semantics, we recall, redefine and develop the notion of \emph{biorder} \cite{Be,cpw}. At its most basic a biorder is  a set with two partial orders, related as follows: 
% We require in addition that the stable order has bounded infima, as these allow the construction of a CCC of stable functions.
\begin{defi}A biorder   $(|D|,\sqleq,\leqs)$  is a set with two partial orders (the \emph{extensional} and \emph{stable} orders) such that  any non-empty set $X \subseteq |D|$ which is bounded above in the stable order (for which we may write $\uparrow X$) has a  greatest lower bound  $\bigwedge X$ in that order, which is also a least upper bound $\bigsqcup X$ for $X$ in the extensional order.   
\end{defi}
\begin{rem}
Note that it is a consequence of this definition that the stable order is contained in the \emph{reverse} of the extensional order, unlike the original definition \cite{Be}, where it appears as a refinement of that order. This is a presentational choice --- it is  possible to reframe the following definitions in the original style --- which has some advantages in the context of interpreting non-deterministic computation:
\begin{itemize}
\item May-and-must duality  --- as we have noted, the may and must orders (which correspond to the extensional order and its dual) are dual to each other in the finite setting of $\Lambda^+$ --- one is a meet semilattice, the other is a join semilattice. Our definition allows a single notion of stability, rather than (in the may testing case) using its dual, co-stability.  
\item Continuity --- introducing fixed points breaks the duality between may and must (and opens the space between finite and unbounded nondeterminism in the must case). Our definition of biorder accommodates this by allowing the definition of stably continuous and extensionally monotone functions (must testing) versus  extensionally continuous and stable functions (may testing) in the same setting.      
\item In our intensional model, which essentially extends sequential algorithms with explicit divergences, it is natural to interpret the extensional order as a \emph{containment} between states, but the stable order as a \emph{refinement} which removes divergences. 
\item  We identify a new property of programs with unbounded nondeterminism --- that they preserve suprema of extensional chains, but only up to stable approximation --- which is most naturally presented as a weak continuity property.
\end{itemize}
\end{rem}
We shall say that a biorder is \emph{extensionally pointed} if it has an extensionally least element, $\ebot$, \emph{stably pointed} if it has a stably least element $\sbot$ (note that the latter element is greatest in the extensional order), and just \emph{pointed} if it has both.

Key constructions on biorders include:
\begin{itemize} 
\item For any family of biorders $\{D_i \ |\ i \in I\}$, the (set-theoretic) product $\Pi_{i \in I}D_i$ and disjoint sum $\coprod_{i \in I}D_i$ with both orders  defined pointwise.
\item If $D$ is a biorder, its stable lifting $D_\up$ is given by adding a new element to $D$ which is least with respect to the stable order (and thus greatest with respect to the extensional order).
\end{itemize}
From these constructions we obtain the empty biorder, the one-element biorder $1$ and its stable lifting --- the (pointed)  biorder $\Sigma$, with two  elements $\{\sbot,\ebot\}$ such that $\ebot \sqleq \sbot$ and $\sbot \leqs \ebot$. For any set $X$, we write  $\widetilde{X}$ for the disjoint sum $\coprod_{x \in X}1$  (in which the extensional and stable orders are both the discrete order).

To define a category of biorders, we recall Berry's notion of \emph{stable function}.
\begin{defi}A function $f:D \rightarrow E$ between partial orders  is \emph{stable} if it is monotone and for any $x \in D$ and $y \leq f(x)$ there exists $m \leq x$ such that $y \leq f(m)$ and for all $z \leq x$, $y \leq f(z)$ implies $m \leq z$.     
\end{defi}
The relationship between stability and bounded infima may be summed up as follows: 
\begin{lem}
If partial orders $D$ and $E$ have bounded infima then $f:D \rightarrow E$ is stable  if and only if it preserves all bounded infima.
\end{lem}
\begin{proof}
From left-to-right, suppose $Z$ is bounded above by $x$, then taking   $y = \bigwedge_{z \in Z}f(z)$   there exists $m$ such that $y \leq f(m)$ and for all $z \in Z$, $m \leq z$ and thus $\bigwedge_{z \in Z}f(z)  = f(\bigwedge Z)$ as required. 

From right-to left, given $y \leq x$, take $m = \bigwedge \{a \leq x\ |\ y \leq f(a)\}$.
\end{proof}
We define a category $\B$ in which objects are biorders, and morphisms from $D$ to $E$ are functions from $|D|$ to $|E|$ which are \emph{monotone stable} ---  i.e. monotone with respect to the extensional order and stable with respect to the stable order.  This concrete category is Cartesian closed --- the product of biorders is a cartesian product and the internal hom $[D,E]$ is the biorder consisting of the set of monotone stable functions from $D$ to $E$ with the extensional (Scott) and stable (Berry) orders on functions --- i.e. 
\begin{center} 
$f \sqleq g$ if for all $x \in D$, $f(x) \sqleq g(x)$\\
$f \leqs g$ if for all $x,y \in D, x \leqs y \Longrightarrow f(x) \leqs g(y)$ and $ f(x) = f(y) \wedge g(x)$. 
\end{center}
The infimum of a stably bounded set of monotone stable functions is given pointwise --- i.e. $(\bigwedge F)(x) = \bigwedge_{f \in F}f(x)$, which is clearly stable and monotone, and a $\sqleq$-supremum for $F$.

  It is well known (and an interesting exercise to prove) that stability of $f:D \times E \rightarrow F$ implies stability of  its currying  $\Lambda(f):D \rightarrow [E,F]$ and vice-versa.  Thus $\B(D \times E, F)$ and $\B(D, [E,F])$ are naturally isomorphic and so $\B$ is Cartesian closed, as is its subcategory of pointed objects. So, in particular, it contains may and must testing denotational semantics of  $\Lambda^+$ in which  the ground type $\o$ is interpreted as the biorder $\Sigma$, the choice operator as the extensional join operator $\sqcup: \Sigma \times \Sigma \rightarrow \Sigma$ (which is monotone stable), and the constants $\mho$ and $\Omega$ as $\sbot$ and $\ebot$ (respectively) in the may testing model and $\ebot$ and $\sbot$ in the must testing model. 
\begin{prop}[Computational Adequacy]$t\Downarrow^\may \mho$ if and only if $\[t\]_\may = \[\mho\]_\may$ and $t\Downarrow^\must \mho$ if and only if $\[t\]_\must = \[\mho\]_\must$  
\end{prop}
Proof is simple based on strong normalization of the $\lambda$-calculus and soundness of the reduction rules. We will give a proof of adequacy for a conservative extension of this model (Proposition \ref{adq}).

\subsection{Sequentiality and Universality}
Sequentiality for higher-order functions  may be defined in  a variety of ways, which will be discussed further in later sections: via correspondence with an explicitly sequential representation  such as games or sequential algorithms (see Section~\ref{sec:s5}) via  Kahn-Plotkin sequentiality indices \cite{KP} 
 (see Section~\ref{sec:s6}), or via universality and full abstraction results for an interpretation of a sequential language (here, for $\Lambda^+$ and in Section~\ref{sec:s7} for a more expressive extension). First, we recall from \cite{ufpc} a  simple result establishing that  conditionally multiplicative functions on  biorders are sequential in the ``first-order'' sense of Milner and Vuillemin. 
\begin{defi}Let $\{A_i\ |\ i \in I\}$ be pointed biorders. A function $f:\Pi_{i \in I}A_i \rightarrow B_\up$ is \emph{strict} if $f(\sbot) = \sbot$ and \emph{$i$-strict} if $\pi_i(x) = \sbot$  implies $f(x) = \sbot$. 
\end{defi}
\begin{prop}[Milner-Vuillemin Sequentiality]
\label{seq}
  Any strict morphism $f:\Pi_{i \in I}A_i \rightarrow B_\up$ is $i$-strict for some $i \in I$. 
\end{prop}
\begin{proof}Define (for $x \in \Pi_{i \in I}A_i$ and $y \in A_i$), the insertion $x[y]_i \in \Pi_{i \in I}A_i$  by $\pi_j(x[y]_i) = y $ if $i =j$ and $\pi_j(x[y]_i) = \pi_i(x)$, otherwise. 

Then $\{\ebot[\sbot]_i \ |\ i \in I\}$ is stably bounded above by $\ebot$, and has stable greatest lower bound $\sbot$ so $f(\bigwedge_{i \in I}\ebot[\sbot]_i\ |\ i \in I\}) = f(\sbot) = \sbot$ and so by stability of $f$, $f(\ebot[\sbot]_i) = \sbot$ for some $i \in I$. Hence  $f(x[\sbot]_i) = \sbot$ for all  $x \in \Pi_{i \in I}A_i$, as $\ebot[\sbot]_i \sqleq x[\sbot]_i$.
\end{proof}
 Note that a function may be $i$-strict for several values of $i$  (for example, the function from $\Sigma \times \Sigma$ to $\Sigma$ sending $x,y$ to $x \sqcup y$). In other words  our model contains  weakly sequential functions, as expected.

It was shown in \cite{ufpc} that any (Milner-Vuillemin) sequential and order-extensional model of Unary PCF in a CCC which interprets the ground type as the two-point order, is  universal (i.e. every element is denoted by a term) and thus fully abstract. We sketch a simplified version of the proof here, which is the basis for a proof of universality for an infinitary extension of $\Lambda^+$ in Section~\ref{sec:s7}.
\begin{prop}\label{univunary}The biorder semantics of $\Lambda^+$ are universal.
\end{prop}
\begin{proof}By duality, we may give the proof for the must testing model only. For any types $S$ and $T$, and $n \in \Na$, write $S \unlhd T^n$ if there is a \emph{definable retraction} from $S$ to $T^n$ -- that is, there are terms $ \inj_1:S \rightarrow T,\ldots ,\inj_n:S \rightarrow T$ and $\proj:T \rightarrow \ldots \rightarrow T \rightarrow S$  such that $\[\lambda x.\proj\spc \inj_1(x)\ldots \inj_n(x)\] = \[\lambda x.x\]$.

The following facts are evident:
\begin{itemize}
\item If $S \unlhd T^n$ then universality at  type $T$ implies universality at type $S$.   
\item Universality holds at type $\o \rightarrow \o$, since the only elements of $[\Sigma,\Sigma]$ are the  functions denoted by  $\lambda x.\Omega$, $\lambda x.x$ and $\lambda x.\mho$. 

\end{itemize}
So it suffices to prove for all types $T$, that $T \unlhd (\o \rightarrow \o)^n$ for some $n$, by induction on $T$, for which the key step  is to show  that $(\o \rightarrow \o)_1 \rightarrow \ldots \rightarrow (\o \rightarrow \o)_n \rightarrow (\o \rightarrow \o) \unlhd (\o \rightarrow \o)^{2^{2n+1}}$, which follows from these two facts:
\begin{itemize}
\item $(\o \rightarrow \o) \rightarrow \o \unlhd (\o \rightarrow \o)^2$ --- by the retraction $\inj_1 \triangleq \lambda f.\lambda x.f \lambda y.x$, $\inj_2 \triangleq \lambda f.\lambda x.f\spc \lambda y.y$ and $\proj \triangleq \lambda x.\lambda y. \lambda z.x\spc (z\spc (y\spc \Omega))$. For each element $e \in \[(\o \rightarrow \o) \rightarrow \o\]$ (i.e. $e \in \{\lambda y. \Omega, \lambda y.\mho,\lambda y.y\spc \Omega,\lambda y.y\spc \mho\}$ we have $(\proj\spc \inj_1(e))\spc \inj_2(e) = e$.
\item $\o \rightarrow \o \rightarrow \o \unlhd (\o \rightarrow \o)^2$  by the retraction $\inj_1 \triangleq \lambda f.\lambda x. (f\spc x)\spc \mho$, $\inj_2 \triangleq \lambda f.\lambda x.(f\spc \mho)\spc x$ and $\proj \triangleq \lambda f.\lambda g.\lambda x . \lambda y. f(x) + g(y)$. 
For any element $e \in \[\o \rightarrow \o \rightarrow \o\]$,\\ $(\proj\spc \inj_1(e))\spc \inj_2(e) = \lambda x.\lambda y.((e \spc x)\spc \tops) \wedge ((e \spc \tops)\spc y) = \lambda x.\lambda y.(e\spc x)\spc y = e$ by stability.  \qedhere
\end{itemize}
\end{proof}
\begin{prop}[Full abstraction] $t \lesssim^\may s$ iff  $\[t\]_\may \sqleq \[s\]_\may$ and $t \lesssim_\must s$ iff $\[s\]_\must \sqleq \[t\]_\must$.
\end{prop}
\begin{proof}
This follows from adequacy, universality and extensionality of the model following standard arguments as in \cite{Plo}.
\end{proof}
 Thus, we have given a fully abstract semantics of a simple, sequential non-deterministic functional language using biorders. In the remainder of the paper, we will extend it to a a fully expressive model of computation, and give an intensional characterization of its types and programs.

\section{\espcf and its semantics}
\label{sec:s3}
$\Lambda^+$ is  too limited to describe generalized higher-order non-deterministic computation because it lacks infinitary datatypes. In particular, this means that there is no real distinction between may and must testing  (witness the formal duality between them) and there is no possibility of describing unbounded nondeterminism and comparing it with the binary choice operation. So we will define a new prototypical language by extending $\Lambda^+$ with a type $\nt$ of natural number \emph{values} --- i.e. terms of type $\nt$ do not have computational effects (nondeterminism or non-termination). Maintaining this distinction between computation and value types (as in \emph{call-by-push-value} \cite{cbpv}) simplifies programming with non-determinism and allows construction of fully-abstract semantics of richer languages by CPS interpretation.  

\emph{Types} of \espcf are either \emph{pointed} (computation) types $P$ or possibly unpointed (value) types  $V$, given by the grammars:

\medskip
\begin{tabular}{l l l l}
$V::=$&$ \nt $&$ | $&$ P$\\
$P::=$&$ \o $&$ | $&$ V \rightarrow P$\\
\end{tabular}
\medskip

\noindent Terms are formed by extending $\Lambda^+$ with the following constants and operations:
\begin{itemize}
\item Expressions of type $\nt$, given by the grammar $e:: =  {\mathtt{0}}\ |\ x:\nt\ |\ \suc(e)$. %Arithmetic constants and operations on expressions of type $\nt$ % ---  we don't specify these but assume that they are sufficient to express the numerals and various other primitive recursive arithmetic functions.
\item Equality testing: given expressions $s,t:\nt$,  $\eq(s,t):\boo$. % reduces to $\tru$ (i.e. $\lambda xy.x$) if its arguments evaluate to the same value, and $\ff$ (i.e. $\lambda x.\lambda y.y$)  if they do not.
\item Fixed point combinators: $\Y_P:(P\rightarrow P) \rightarrow P$ for each pointed type $P$.
\item Unbounded Choice:  ${?_N}:(\nt \rightarrow \o) \rightarrow \o $, which nondeterministically chooses a numeral to  which its argument is applied. 
\end{itemize}
\espcf may be regarded as a target language for \emph{continuation-passing-style} translation of languages with bounded and unbounded nondeterminism, with $\o$ as the return or answer type. Most simply, by defining the type $\nat$  of natural number computations to be the type  $(\nt \rightarrow \o) \rightarrow \o$, we may express the language  SPCF   (observably sequential PCF \cite{CF}) extended with bounded  (binary) or unbounded (natural number) choice. In particular, note that we may define $\pred:\nt \rightarrow \nat \triangleq \lambda u.\lambda f.(\Y\lambda g.\lambda v.(\eq(u,\suc(v))\spc (f\spc v))\spc (g\spc \suc(v)))\spc {\mathtt{0}}$   

 SPCF is itself an extension of PCF with a simple non-local control operator (e.g. we may define $\callcc:((\nat \rightarrow \nat) \rightarrow \nat) \rightarrow \nat \triangleq \lambda f.\lambda k.(f \spc \lambda xy.x\spc k)\spc k$) and some error terms which immediately abort computation (in this case, a single such term, $\e$). SPCF provides  a complete syntactic representation of the (strongly) observably sequential  functions (and the corresponding deterministic sequential algorithms)  \cite{CCF,bist} as demonstrated by full abstraction and universality results --- this paper may be seen as an extension of these results to weakly observably sequential functions and non-deterministic sequential algorithms.

\subsection{Operational Semantics}
We give may and must testing operational semantics for programs (closed terms of type $\o$) by extending  the  rules for  $\Lambda^+$ (Table \ref{opsem1}) with the additional rules in Table \ref{opsem}.
 \begin{table}
\begin{center}
\begin{tabular}{c c c} 
 {\Large$ \over \eq({\mathtt{n}},{\mathtt{n}})\Downarrow \lambda x.\lambda y.x$} & &  {\Large$ \over \eq({\mathtt{m}},{\mathtt{n}})\Downarrow \lambda x.\lambda y.y$}\mbox{$m \not = n$} \\\\
\end{tabular}
\begin{tabular}{c c c c c}
 {\Large $\over \Y  \Downarrow \lambda f.f\spc (\Y f)$}& &{\Large $\exists n. t\spc {\mathtt{n}} \Downarrow^\may \mho \over  ?_N\spc t \Downarrow^\may \mho$} & & {\Large $\forall n. t\spc {\mathtt{n}} \Downarrow_\must \mho \over  ?_N\spc t \Downarrow_\must \mho$}\\\\
\end{tabular}
\caption{Operational Semantics for \espcf}\label{opsem}
\end{center}
\end{table}
The definitions of  may, must and may-and-must approximation and equivalence extend directly from $\Lambda^+$. Also as in $\Lambda^+$, the presence of the error element makes the language extensional --- any terms of function type are observationally  equivalent if (and only if) applying to the same argument returns  equivalent results. (This is essentially Milner's Context Lemma  \cite{mill} and may be proved using the same techniques.)  In particular, we will use the result for must equivalence in Section~\ref{sec:s7}.
\begin{lem}[Function Extensionality]\label{ext}For any closed  terms $s,t:S \rightarrow T$, if $s\spc r \simeq_{\must} t\spc r$ for all $r:S$ then $s \simeq_{\must} t$.     
\end{lem}
\subsection{Unbounded Nondeterminism}
\espcf gives us a setting in which we may express programs with \emph{unbounded} non-determinism --- they may evaluate to  (countably)
infinitely many different values without diverging.  Evidently, we may express  bounded choice  up to may-and-must equivalence using unbounded choice --- e.g. $s \spc\orr\spc t \simeq^\may_\must   ?_N\spc (\lambda k. (\eq(k,0)\spc s)\spc  t)$. Conversely:
\begin{prop}$?_N$ is macro-expressible up to may equivalence in bounded \espcf.
\end{prop}
\begin{proof}The terms 
 $?_N$  and   $\lambda k.(\Y_{\nat} \lambda f.\lambda u.(k\spc u)  \orr\spc (f\spc (\suc\spc x)))\spc {\mathtt{0}}$  are may-equivalent. (They denote the same elements in our fully abstract model of may testing.)
\end{proof}
Note that this equivalence fails with respect to must equivalence (evaluation may always take the right hand branch of the binary choice and so diverge), as will any such attempt to define countable choice using bounded choice  (a term which reduces to infinitely many different values will have an infinite reduction path by K\"onig's lemma and may therefore diverge). In other words:       
\begin{prop}\label{cnd}$?_N$ is \emph{not} macro-expressible up to must equivalence using binary choice. 
\end{prop}
Formal proof of this claim is straightforward using our fully abstract model (Section~\ref{sec:s7}).
\begin{rem}[Fairness]
One reason for studying unbounded nondeterminism is its close connection with the notion of \emph{fairness}, according to which an event will eventually occur, even if it is not possible to bound the number of steps taken before this happens. For example, suppose we wish to implement a type $\coin$ with constants 
$\newcoin:(\coin \rightarrow \o) \rightarrow \o$ (which  supplies  a coin to its argument) and 
$\toss: \coin \rightarrow (\coin \rightarrow \o) \rightarrow (\coin \rightarrow  \o) \rightarrow \o$ which tosses its first argument and chooses its second or third argument depending on whether the result is heads or tails, and passes the tossed coin to it. 

A necessary condition for an implementation of $\coin$  to be \emph{fair} is that tossing a new coin will eventually come up heads, but may require an unbounded number of tosses to do so. In other words, a program which repeatedly  tosses a coin and increments a counter until it gets a head, and then returns the contents of the counter may  return any number value (but not diverge) 
 --- i.e. $$\lambda k.((\Y_{\nt \rightarrow \coin \rightarrow \o}\lambda f.\lambda u:\nt.\lambda x:\coin.(\toss\spc x)\spc \lambda y.(k\spc u) \spc (f\spc \suc(u)))\spc 0)\spc \newcoin$$ is must equivalent to  $?_N$.    So a fair coin may be used to express unbounded choice and is therefore not definable in bounded \espcf by Proposition \ref{cnd}. Conversely, in unbounded \espcf we may implement a coin which passes the fairness test, e.g. by defining the macros $\coin \triangleq \nt$, $\newcoin \triangleq ?_N$ and $\toss \triangleq \lambda u.\lambda f.\lambda g.(\eq(u,0)\spc (?_N\spc f))\spc (\pred(u)\spc g)$.
\end{rem}

\begin{rem}[Abstract Interpretation]Another setting in which unbounded nondeterminism may arise is in the ``abstract interpretation'' of a domain with infinitely many states as a finite domain --- the ``concretization'' of such an abstract interpretation may then represent a single abstract state as an unbounded  choice of concrete states.  Observe that in \espcf, countable nondeterminism may be used to define an embedding-projection pair from 
the type $\nt \rightarrow \o$ to $\o$ --- the terms $e \triangleq \lambda x.\lambda u.x:\o \rightarrow \nt \rightarrow \o$ and
$p \triangleq ?_N:(\nt \rightarrow \o ) \rightarrow \o$ satisfy $\lambda x. p\spc (e\spc x) \simeq_\must \lambda x.x$ and $\lambda y.e\spc (p\spc y) \lesssim_\must \lambda y.y$. In other words, there is a Galois insertion between $\nt \rightarrow \o$ and $\o$, giving an \emph{abstract interpretation} of the former, infinitary type inside the latter, finitary one, which we may use to define, inductively, a  Galois connection between each pointed type of \espcf  and the corresponding type of $\Lambda^+$ obtained by erasing all instances of $\nt$.   
 \end{rem}

\subsection{Continuity}
A challenge encountered in defining and reasoning about an adequate denotational semantics of  must testing for programs with unbounded non-determinism is that it cannot be continuous with respect to the must ordering (on must equivalence classes of terms). 
 
For instance, define $t_n:\nt \rightarrow \o \triangleq \Y_{\nt \rightarrow \o}\lambda f.\lambda u.(\eq({\mathtt{n}},u)\spc \e) \spc (f\spc \suc(u))$, so that $t_n\spc {\mathtt{i}}\Downarrow_\must$ iff $i \leq n$.  It is easy to see (and prove using our fully abstract model) that $t_n  \lesssim_{\must} t_{n+1}$ for each $n \in \Na$, and that the  $\lesssim_\must$-least upper bound for this chain of terms (up to must equivalence) is the term $\lambda f.\e$. However,  whereas $?_N \spc t_n$ may diverge for all $n$,  $?_N\spc (\lambda f.\e)$ must converge --- i.e. $?_N$ cannot denote a function which is continuous with respect to the must ordering.

Another example shows that function application is not continuous in the function component either --- or in other words, that least upper bounds of chains of functions are not given pointwise, in general. 
 For instance,  let $s_k:(\nt \rightarrow \o)\rightarrow \o  \triangleq \lambda f.?_N \lambda x.f\spc \suc^k(x)$, which nondeterministically supplies to its argument a numeral greater than or equal to $k$. It is evident that $s_k  \lesssim_{\must} s_{k+1}$ for each $k \in \Na$, and that the $\lesssim_\must$-least upper bound for this chain of terms (up to must equivalence) is the term $\lambda f.\e$. However,  $s_k \spc \lambda u.\Omega$ diverges for all $k$, but  $(\lambda f.\e)\spc \lambda u.\Omega$ must converge.

This suggests that we cannot define \emph{well-behaved} fixed points for functions by taking least fixed points of chains of approximants with respect to the must approximation ordering. This problem will be resolved semantically by defining least fixed points with respect to the stable order  rather than the extensional (observational) order. Application  is stably-continuous with respect to chains of functions, although not their arguments. (The $t_i$ defined above denote a stable chain in our model, but the $s_k$ do not.)  

Failure of continuity also entails that there are must inequivalent terms which cannot be distinguished by finitary tests (i.e. by applying them to arguments which are compact in the must ordering). We give an example at the end of this section, after developing the necessary interpretation of fixed points.

\subsection{Complete Biorders}
To extend our model of $\Lambda^+$ to  \espcf we  require biorders which are directed complete with respect to at least one of their orders, in order to define least fixed points with respect to that order. We now define these notions of complete biorder.
\begin{defi}An $I$-indexed family $x$ of elements of a biorder $D$ is $\sqleq$-directed if for all $i,j \in I$ there exists $k \in  I$ such that $x_i,x_j \sqleq x_k$. $D$ is \emph{extensionally  complete} if the extensional order is directed  complete --- i.e. every such family has a least upper bound, and for any $\sqleq$-directed $I$-indexed families $x$ and $y$, $x_i \sqleq y_i$ for all $i\in I$ implies $\bigsqcup_{i \in I} x_i \leqs \bigsqcup_{i \in I}y_i$. 
\end{defi}
 It is not hard to show that the product, disjoint sum, lifting and function-space preserve extensional completeness.

The requirements for a biorder to be \emph{stably complete} are slightly more involved. They include a ``bounded distributivity condition'' requiring certain bounded meets to distribute over directed joins, which will be used  to define CCCs of  stably complete biorders.
\begin{defi}A \emph{stable upper bound} for a set $X$ of  stably directed, $I$-indexed families over a biorder $D$ is an $I$-indexed, stably directed family $y$ such that for all $x \in X$ and $i \in I$, $x_i \leqs y_i$, and for all $i,j \in I$, $x_i \wedge y_j = x_j \wedge y_i$.  

 $D$  is \emph{stably complete} if the stable order is  directed complete  and for any stably bounded set $X$ of stably directed $I$-indexed families of elements,   $\bigwedge_{x \in X}\bigvee_{i \in I}x_i = \bigvee_{i \in I}\bigwedge_{x \in X}x_i$.   

A biorder is \emph{bicomplete} if it is both extensionally and stably complete.
\end{defi}
 It is easy to see that the coproduct and product of families of stably complete domains are stably complete. To show that  lifting preserves stable completeness  (so in particular $\Sigma$ is stably complete) is a little harder. Note  that the above distributivity property does not hold in general for unbounded sets of stably directed families over $\Sigma$ (i.e. it is not a completely distributive lattice) --- take, for example the set of families $\{\{n_i\ |\ i \in \omega\} \ |\ n \in \Na\}$ with $n_i = \tops$ if $i > n$ and $n_i = \sbot$ otherwise, which satisfies $\bigvee_{i \in \omega}\bigwedge_{n \in \Na}n_i = \sbot$ and $\bigwedge_{n \in \Na}\bigvee_{i \in \omega}n_i = \tops$.    
\begin{lem}If $D$ is stably complete then $D_\up$ is stably complete.
\end{lem} 
\begin{proof}
We check that the bounded distributivity property holds. 
Suppose the set $X$ of $I$-indexed directed families is bounded above by $\{y_i \ |\ i \in I\}$. By definition, $\bigvee_{i \in I}\bigwedge_{x \in X}x_i \leqs \bigwedge_{x \in X}\bigvee_{i \in I}x_i$. If $\bigvee_{i \in I}\bigwedge_{x \in X}x_i \not= \sbot$ then defining $J = \{i \in I\ |\ \bigwedge_{x \in X}x_i \not = \bot_S$, we have a set of  $J$-indexed directed sets of elements of $D$,    bounded above by $\{y_j\ | \ j \in J\}$ and so $\bigvee_{i \in I}\bigwedge_{x \in X}x_i = \bigvee_{j \in J}\bigwedge_{x \in X}x_j = \bigwedge_{x \in X}\bigvee_{j \in J}x_j = \bigwedge_{x \in X}\bigvee_{i \in I}x_i$ by stable completeness of $D$. 
 
So we  need to consider the case in which  (a)  $\bigvee_{i \in I}\bigwedge_{x \in X}x_i = \sbot$ and (b) $\bigwedge_{x \in X}\bigvee_{i \in I}x_i \not = \sbot$. We claim that this implies $x_i = \sbot$ for all $i \in I$ and $x \in X$ and so in fact $\bigwedge_{x \in X}\bigvee_{i \in I}x_i = \sbot$ as required.
To prove the claim, choose any $i \in I$. There exists $z \in X$ such that $z_i = \sbot$ by $(a)$ and $j \in I$ such that $z_j \not = \sbot$ by $(b)$. Because $z_i \wedge y_j  = y_i \wedge z_j$, we have   $y_i = \sbot$, and hence  $x_i = \sbot$ for all $x \in X$ as required.
\end{proof}
 %We will construct Cartesian closed categories of stably complete and bicomplete biorders, giving further examples. 
\begin{lem}\label{la}If  $E$ is stably  complete then $[D,E]$  is stably complete.
\end{lem}
\begin{proof}
Joins of stably directed sets are defined pointwise (i.e. function application is stably continuous with respect to functions). To show that these are stable, we need to show that they commute appropriately with bounded meets using the bounded distributivity property. 
Suppose $F$ is a stably directed set of monotone stable functions from $D$ to $E$, and $X$ is a set of elements stably bounded above by $y$. Then the set $\{\{f(x) \ |\ f \in F\} \ |\ x \in X\}$  of stably directed $F$-indexed sets is stably bounded above by $\{f(y)\ |\ f \in F\}$ --- for any $x$, we have $f(x) \leqs f(y)$ for all $f$ and for any $g \in F$ there exists $h \in F$ with $f,g \leqs h$  and so $f(x)  \wedge g(y) = f(y) \wedge h(x) \wedge g(y)  = f(y) \wedge g(x)$. Thus, by the bounded distributivity condition:
 $(\bigvee F)(\bigwedge X) = \bigvee_{f \in F}\bigwedge_{x \in X}f(x) = \bigwedge_{x \in X}\bigvee_{f \in F}f(x) = \bigwedge_{x \in X}(\bigvee F)(x)$.
\end{proof}
 So, in particular, the full subcategory of $\B$ consisting of bicomplete biorders is Cartesian closed. We define the  following subcategories of $\B$ in which morphisms are continuous:  
\begin{itemize}
\item $\EB$ --- objects are extensionally complete biorders and morphisms from $D$ to $E$ are functions from $D$ to $E$  which are continuously stable  --- i.e. they are monotone stable and preserve  suprema of extensionally directed sets.
\item $\SB$ --- objects are stably  complete  biorders and morphisms from $D$ to $E$ are  functions from $D$ to $E$ which are  stably continuous --- i.e. they are monotone stable and preserve  suprema of stably directed sets.
\end{itemize}
Each of these categories is Cartesian closed, with products given pointwise, and
internal homs given the by relevant biorders of functions --- i.e. $[D,E]_{\EB}$ is the restriction of $[D,E]$ to continuously stable functions and $[D,E]_{\SB}$ is the restriction of $[D,E]$ to stably continuous functions. We need to check that in each case this defines an object of our category --- i.e. is  (respectively) extensionally complete or stably complete.  We give the second (more interesting) case.
\begin{lem}If $D$ and $E$ are stably complete then $[D,E]_{\SB}$  is stably complete.
\end{lem}
\begin{proof}
The proof follows that of Lemma~\ref{la}. We also need to show that if $F$ is a set of continuously stable functions, stably bounded above by a (continuous) function $g$, then $\bigwedge F$ is continuous:
 if  $X$ is a  stably directed set then $\{\{f(x)\ |\ x \in X\}\ |\ f \in F\}$ is stably bounded above by the stably directed set $\{g(x)\ |\ x \in X\}$ since for all $x,x' \in X$ there exists $x'' \in X$ with $x,x' \leqs x''$ and thus $f(x) \wedge g(x') = f(x'') \wedge g(x) \wedge g(x') = f(x'') \wedge g(x') \wedge g(x) = f(x') \wedge g(x)$.   \\
Hence $(\bigwedge\! F)(\bigvee\! X) = \bigwedge_{f \in F}f(\bigvee\! X) = \bigwedge_{ f\in F}\!\bigvee_{x \in X}\!f(x) = \bigvee_{x \in X}\!\bigwedge_{f \in F}\!f(x) = \bigvee_{x \in X}\!(\bigwedge \!F)(x)$.    
\end{proof}
Finally, we need to check that the isomorphism of biorders $\Lambda:[D \times E,F] \rightarrow [D, [E,F]]$ restricts to each of our categories --- i.e. $\Lambda(f)$ is respectively  continuously stable or stably continuous  if and only if $f$ is continuously stable/stably continuous.  This is straightforward, based on these properties for the Scott and Berry orders. Thus we have shown:
\begin{prop}The categories $\EB$ and $\SB$  are Cartesian closed.
\end{prop}
%Moreover, we may observe that  the  above argument (Lemma~\ref{la}) also shows that if $E$ is stably (resp. extensionally) complete, then the biorder $[D,E]$ of monotone stable functions is stably (resp.  extensionally) complete, with suprema of directed sets given pointwise in each case 

As an example  we show that for any set $X$, the function space $[\Sigma^X,\Sigma] \cong [[\widetilde{X},\Sigma],\Sigma]$ corresponds to a powerdomain biorder on $X$, justifying the identification of the \espcf type $(\nt \rightarrow \o) \rightarrow \o$ with the type $\nat$ of natural number computations. 
\begin{defi}Given a set $X$, let $\xp(X)$ be the biorder consisting of subsets of $X$, with the  extensional order  being inclusion and the stable order discrete. 
\end{defi} 
There is an evident strict morphism $\phi:\xp(X)_\up \rightarrow [\Sigma^X,\Sigma]$ sending $Y \subseteq X$ to $\bigsqcup_{y \in Y}\pi_y$. To show that this is an isomorphism by giving an inverse, we define:
\begin{defi}
Given monotone stable $f: \Pi_{i \in I}A_i \rightarrow \Sigma$ let $\si(f)$ be the set of $i \in I$ such that $f$ is $i$-strict (the sequentiality indices of $f$). By Proposition \ref{seq} this is non-empty if and only if $f$ is strict.
\end{defi}

\begin{lem}\label{si}If  $\pi_i(x) = \ebot$ for all $i \in \si(f)$ then $f(x) = \ebot$. 
\end{lem}
\begin{proof}If $f$ is not strict then $f(x) \sqleq f(\bot_S) = \top_E$. Otherwise, noting that $i \not \in \si(f)$ implies $f(\ebot[\sbot]_i) = \ebot$, we have  $f(x) \sqleq f(\bigwedge_{i \not\in \si(f)}\ebot[\sbot]_i) = \bigsqcup_{i \not\in \si(f)}f(\ebot[\sbot]_i) = \ebot$ by stability of $f$. 
\end{proof}

\begin{prop}\label{pd} The biorders $\xp(X)_\up$ and $[\Sigma^X, \Sigma]$ are isomorphic in $\B$.
\end{prop}
\begin{proof}
By Lemma~\ref{si}, $f(x) = \sbot$ if and only if $\pi_i(x) \not = \bot_E$  for some $i \in \si(f)$, and thus we may define an inverse to $\phi:\xp(X)_\up \rightarrow [\Sigma^X,\Sigma]$ by $\phi^{-1}(f) = \sbot$, if $f = \sbot$ and $\phi^{-1}(f) = \si(f)$, otherwise.    
\end{proof}  
$\phi$ is continuously stable (so $\xp(X)_{\up} \cong [\Sigma^X,\Sigma]_{\EB}$) but not stably continuous (cf the non-continuity of unbounded choice with respect to must testing). 
\begin{lem}If $f$ is stably continuous then $\si(f)$ is finite.
\end{lem}
\begin{proof}Given $J \subseteq I$ define $a_J \in \Pi_{i \in I}A_i$ by $\pi_i(a_J) = \ebot$ if $i \in J$ and $\pi(a_J) = \sbot$ otherwise. Then $\bigvee_{J \subseteq_{\mathit{fin}} I}a_J =  \ebot$ and so if $f$ is stably continuous, $f(a_J) = \ebot$ for some finite $J \subseteq I$. Then for all $i \in \si(f)$, $i$-strictness of $f$ entails that $\pi_i(a_J) = \ebot$. Thus $\si(f) \subseteq J$ and thus  $\si(f)$ is finite.      
\end{proof}
So defining $\xp_{\mathit fin}(X)$ to be the restriction of $\xp(X)$ to finite subsets of $X$, we have: $\xp_{\mathit fin}(X)_\up \cong [\Sigma^X,\Sigma]_{\SB}$. 
%$(\xp_{\mathit{fin}}(X) \cup \{\sbot\},\subseteq, \{(\sbot,y)\ |\ y \in  \xp(X) \cup \{\sbot\})$.
\subsection{Denotational Semantics of \espcf}
To interpret recursive higher-order programs, we  require fixed points of endomorphisms $f:D \rightarrow D$, where $D$ is  extensionally or stably complete and pointed. These may be obtained in standard fashion, as (respectively, extensional or stable) suprema of the chain of approximants $\bot, f(\bot),\ldots, f^i(\bot), \ldots$. 
In the categories $\EB$ and $\SB$, which are cpo-enriched, the fixed point is reached at $f^\omega$, whereas in the category of biorders and monotone stable functions it may be necessary to continue the chain beyond $\omega$, but it will still reach a stationary point \cite{AP}.
\begin{prop}If $D$ is stably (resp. extensionally) complete and pointed then every  monotone stable  function $f:D \rightarrow D$ has $\leqs$-least (resp. $\sqleq$-least) fixed points. 
\end{prop} 
\begin{proof}
Define the chain of stable approximants  $f^\lambda \in D$ for each ordinal $\lambda$ by:
\begin{itemize}
\item  $f^0 = \sbot$,
\item $f^{\kappa +1} = f(f^\kappa)$, 
\item  $f^\lambda = \bigvee_{\kappa < \lambda} f^\kappa$ if $\lambda = \bigcup_{\kappa <\lambda}\kappa$. 
\end{itemize}
By  Hartog's Lemma  this has a stationary point, which is a $\leqs$-least fixed point for $f$ --- evidently if $f$ is  $\lambda$-continuous (preserves suprema of $\lambda$-chains) then this stationary point is~$f^\lambda$.   
\end{proof}

We now have the basis for may testing and must testing semantics of \espcf in our categories of biorders --- more precisely:
\begin{itemize}
\item May testing and must testing semantics of unbounded \espcf in the category of bicomplete biorders and monotone stable functions. (As we have noted, unbounded choice is expressible up to may-testing  in bounded \espcf, so the bounded/unbounded distinction is moot in this case and we focus on must testing for unbounded nondeterminism).  
\item A must testing semantics of bounded \espcf in the category of stably complete  and stably continuous functions.
\item A may testing semantics of bounded/unbounded \espcf in the category of  extensionally complete and continuously stable functions.
\end{itemize}
(Pointed) types are interpreted as (pointed) biorders : $\o$ denotes the biorder $\Sigma$,  the type $\nt$ denotes the discrete biorder of natural numbers $\widetilde{\Na}$ and $S \rightarrow T$ the relevant function-space.
Terms $x_1:S_1,\ldots,x_n:S_n\vdash t:T$ denote morphisms  from $\[S_1\]\times \ldots \times \[S_n\]$ to $\[T\]$, derived from the Cartesian closed structure of each category, together with:
\begin{itemize} 
\item The extensional join operation: $s \orr t$ denotes the bounded (binary) join $\[s\] \sqcup \[t\]$ and $?_N$ denotes the infinite join  $\bigsqcup_{n \in \Na}\lambda f. f(n)$.  
\item Error: In the must testing models, $\e$ denotes the extensionally least element $\ebot$. In the may testing model, it denotes the stably least element (greatest extensional element) $\sbot$.  
\item Fixed points:  $\Y_P:(P \rightarrow P) \rightarrow P$ denotes a  least fixed point  of the function $Y  \triangleq F \vdash \lambda f.f\spc (F f)$. In the must testing models it denotes the stably least fixed point and in the may testing models it denotes the extensionally least fixed point.
\end{itemize}

\subsection{Computational Adequacy}
Soundness of the reduction  rules with respect to the denotational semantics follows from the structure identified in our categories, yielding: 
\begin{prop}\label{sound}$t \Downarrow^\may C$ implies  $\[t\]_{\may} = \[C\]_{\may}$ and $t \Downarrow^\must C$ implies  $\[t\]_{\must} = \[C\]_{\must}$.   
\end{prop}
We  establish the  converse for terms of type $\o$ (computational adequacy) by defining ``approximation relations'' in the style of Plotkin \cite{fpc}. The models in which this proof is not completely standard are those which are not $\omega$-cpo-enriched --- in particular, the case of  unbounded nondeterminism with must testing, which we give here. Plotkin's proof still adapts to this case because fixed point combinators denote least upper bounds of stable chains, which are defined pointwise at function type.

For each type $T$ we define a relation $\lhd_T$  between elements of $\[T\]$ and closed terms  of type $T$:
\begin{itemize}
\item $e \lhd_\nt t$ if   $e= \[t\]$.
\item $e \lhd_{\o} t$ if $e = \[\mho\]$ implies  $t \Downarrow^\must \mho$. 
\item $f \lhd_{S \rightarrow P} t$ if  $e \lhd_S s$ implies $f(e) \lhd_P t\spc s$. 
\end{itemize}
We want to show that every program is approximated by its denotation, for which we use the following two lemmas, proved by straightforward induction on type structure.
\begin{lem}\label{red} If $t \Downarrow^\must C$ implies $s \Downarrow^\must C$ then $e \lhd_T t$ implies $e \lhd_T s$.
\end{lem}
\begin{lem}\label{l3}
 If $\langle e_\alpha \ |\ \alpha <\lambda\rangle$ is a stable chain of functions such that $e_\alpha \lhd_T t$ for all $\alpha < \lambda$ then $\bigvee_{\alpha< \lambda}e_\alpha \lhd_{T} t$. 
\end{lem}
The key lemma for our proof establishes that the fixed point combinator is approximated by its denotation.
\begin{lem}\label{Y} For any pointed type $P$, $\[\Y_P\] \lhd_{(P \rightarrow P) \rightarrow P}  \Y_P$.
\end{lem}
\begin{proof}
We show by induction on $\lambda$ that  $Y^\lambda \lhd_{(P \rightarrow P) \rightarrow P} \Y$ for each approximant $Y^\lambda$ to $\[\Y\]$ --- i.e. if, $g \lhd_{P \rightarrow P} t$ then $(Y^\lambda g)  \lhd_P \Y t$.
\begin{itemize}
\item For $\lambda = 0$, we have $(Y^\lambda g)  = \sbot \lhd_P \Y\spc t$.
\item If $\lambda = \kappa +1$ then by hypothesis $Y^\kappa \unlhd \Y$, and so  $Y^{\lambda}\spc g  = g\spc (Y^\kappa g) \unlhd_P t\spc (\Y t)$. Since $\Y t \Downarrow^\must C$ implies $t\spc (\Y t) \Downarrow^\must C$, $Y^\lambda\spc g \lhd_P \Y\spc t$  by Lemma~\ref{red}.
\item For $\lambda = \bigcup_{\kappa < \lambda} \kappa$, we have $Y^\lambda = \bigvee_{\kappa <\lambda}Y^\kappa \unlhd_P \[\Y\]$ by  Lemma~\ref{l3}.
\end{itemize}
So $\[\Y\] \lhd_{(P \rightarrow P) \rightarrow P} \Y$ as required.
\end{proof}
The rest of the proof is now standard --- the approximation relation is extended to open terms by defining $f:\[\Gamma\] \rightarrow \[T\] \lhd_{\Gamma,T} \Gamma \vdash t:T$ if $\Gamma = x_1:S_1,\ldots,x_n:S_n$ and $e_1 \lhd_{S_1} s_1,\ldots,  e_n \lhd_{S_n} s_n$ implies  $f(e_1,\ldots, e_n) \lhd_T t[s_1/x_1,\ldots, s_n/x_n]$.
\begin{prop} If $\Gamma \vdash t:T$ then $\[\Gamma \vdash t:T\] \lhd_{\Gamma,T} t$ 
\end{prop}
\begin{proof}
By structural induction on $t$.
\begin{itemize}  
\item  If $t$ is a value  or application then $\[t\] \lhd_{\Gamma,N} t$ by definition of the approximation relation
\item If $t = \lambda x:S.t'$ and $e \lhd_S s$, then $\[t\] \unlhd t'[s/x]$ by hypothesis, and $t'[s/x]\Downarrow^\must C$ implies $\lambda x.t\Downarrow^\must C$ and so $\[t\](e) \unlhd t\spc s$ by Lemma~\ref{red} above.  
\item If  $t = ?_N$, then for any  $e \lhd_{\nt \rightarrow \o} s$, either  $\[s\](e) = \sbot$ or $e(i) = \tops$ for all $i \in \omega$ and so by hypothesis $s\spc i \Downarrow^\must \mho$ for all $i$ and hence $t\spc s \Downarrow^\must$ as required.  
\item If $t = \eq(s,s')$ then $\[t\] \lhd_{o \rightarrow \o \rightarrow \o} t$ by definition.       
\item The case $t = \Y_T$ is dealt with in Lemma~\ref{Y} above.   \qedhere
\end{itemize}
\end{proof}
Thus, by definition of the approximation relation.
\begin{prop}\label{adq}For any program $t:\o$, $\[t\]_\must = \[\mho\]_\must$ implies $t \Downarrow_\must \mho$.
\end{prop}
By a standard argument, computational adequacy for our models of bounded and unbounded \espcf implies inequational soundness.
\begin{cor}If $\[s\]_\must \sqleq \[t\]_\must$ then $t \lesssim_\must s$, and if $\[t\]_\may \sqleq \[s\]_\may$ then $t \lesssim_\may s$.
\end{cor} 
 As promised, we may now give an example of must inequivalent terms which cannot  be separated by any finite approximant to the argument which does separate them. Let $U = \nt \rightarrow \o \rightarrow \o$, so that an element $e \in \[U\]$ is compact with respect to must approximation if and only if there exists $j$ such that $\[e\](k) = \bot_S$ for all $k > j$. 

\begin{exa}\label{ncs}
Consider the terms:
\begin{itemize}
\item  $s:U \rightarrow \o \triangleq \lambda f.(\Y_{\nt \rightarrow \o}\lambda F.\lambda u. (f\spc u)\spc (F\spc (\suc(u))))\spc {\mathtt{0}}$    
\item  $t:U \rightarrow \o \triangleq \lambda f. ?_N\spc (\lambda v.(\Y_{\nt \rightarrow \o}\lambda F.\lambda u. (\eq(u,v)\spc \mho)\spc ((f\spc u)\spc (F\spc (\suc(u)))))\spc {\mathtt{0}})$.
\end{itemize}
Then $\[s\]$ is the least upper bound of the stable chain of approximants  $\langle \[s_i\] \ |\  i \in \omega\rangle$, where $s_i = (f 0)\spc ((f 1)\spc (\ldots ((f i)\spc \Omega)\ldots ))$ and $\[t\]$ is the greatest lower  bound of the descending extensional chain  $\langle \[t_i\]\ |\ i \in \omega \rangle$, where $t_i = (f 0)\spc ((f 1)\spc (\ldots ((f i)\spc \mho)\ldots ))$. 
Then:
\begin{itemize}
\item $\[s_i\]_{\must} \sqleq \[t_i\]_{\must}$ for all $i \in \omega$, so $\[s\]_{\must} \sqleq \[t\]$ and thus $s \lesssim_{\must} t$.
\item $s$ and $t$ are distinguished by  application to $\lambda u.\lambda x.x$, since we have $s\spc \lambda u.\lambda x.x \not \Downarrow^\must \mho$ but      $t\spc \lambda u.\lambda x.x \Downarrow^\must \mho$. 
\item If $e:U$ is a finite $\lmust$-approximant to $\lambda u.\lambda x.x$ (i.e. $\lmust$-dominates finitely many elements, which is equivalent to compactness)
   then $s\spc e \Downarrow^\must$ if and only if $t\spc e \Downarrow^\must$, since there exists   $j$ such that $\[e\](k) = \bot$ for all $k > j$ and thus $s_i\spc e \Downarrow^\must$ if and only if  $t_i\spc e \Downarrow^\must$ for all $i > j$. 

\end{itemize}
\end{exa}

We will study the completeness (full abstraction) properties of our models in Section~\ref{sec:s7}. First, we develop a setting in which we can give intensional characterizations of them.

\section{Ordered Concrete Data Structures}
\label{sec:s4}
We will now define the notion of ordered concrete data structure and use it to give an intensional representation of monotone stable functions. It is based on the  original notion of concrete data structure, introduced by Kahn and Plotkin \cite{KP} and further developed by Berry and Curien \cite{BeCu}.

A concrete data structure  consists of  sets of \emph{cells}, \emph{values} and \emph{events} (which are pairs of cells and values), and an \emph{enabling relation} between events and cells. The idea is that each step of a  sequential computation is represented as an event (the filling of a cell with a value), which may be dependent on some combination of previous events having occurred (as specified by the enabling relation).  Programs correspond to \emph{states} which specify values for filling enabled cells --- sets of events which satisfy two conditions: \emph{consistency} --- every cell must be filled with a unique value --- and \emph{safety} --- for every filled cell there is a finite chain of enablings of filled cells within the state, back to an ``initial cell'' which does not depend on any prior events. In order to model nondeterministic computation  we adapt this setting in the following ways:
\begin{itemize} 
\item  Removing the consistency condition, so that a cell may be filled with multiple different values. 
\item  Placing an ordering on cells and values, (and  thus events) and requiring states to be upwards closed under this ordering. This reflects the fact that (for example) the responses of a program to an input which is a nondeterministic choice of $a$ and $b$ must include all of its responses  to both $a$ and $b$. 
\item  Including a distinct element $\bullet$ --- representing failure by divergence or error, depending on may or must testing interpretation --- with which  any cell may be filled (cf the representation of divergence in the game semantics of must testing in \cite{HM}). 
\item Extending the safety condition to allow \emph{transfinite} ordinal  chains of enabling events. (Capturing the distinction between infinite, and finite but unbounded interaction.) 
\end{itemize}

\begin{rem}We may think of a concrete data structure as a kind of \emph{two-player game} in which one player chooses a cell for the second player to fill with a value of their choice, enabling a choice of cells for the first player, and so on. 
In the deterministic setting, this analogy can be made into a precise correspondence between the category of concrete data structures and sequential algorithms, and a category of \emph{graph games} and strategies \cite{HyS}. However,  this  correspondence depends on determinacy to decompose the CCC of sequential algorithms into a model of linear logic from which the graph games model is constructed: we leave the description of a more explicitly games-like characterization of ordered concrete data structures as future work.
\end{rem}

\subsection{Ordered Concrete Data Structures}
\begin{defi}A (filiform) \emph{ordered concrete data structure} (ocds) $A$ is given by a tuple $(C(A),V\!(A),\vdash_A,E(A))$ where $C(A),V\!(A)$ (the \emph{cells} and \emph{values} of $A$)  are  partial orders not containing the distinguished element $\bullet$, $E(A) \subseteq C(A) \times V\!(A)$ is a set of \emph{events} and  $\vdash_A \subseteq (E(A) \cup \{*\}) \times C(A) $ is a relation (\emph{enabling}) such that $(c,v) \vdash c'$ implies $c < c'$. 
\end{defi}
 We write $V\!(A)_\bullet$ for the partial order on $V\!(A) \cup \{\bullet\}$ with $a \leq a'$ if $a = \bullet$ or $a \leq a'$ in $V\!(A)$.\\
We write  $E(A)_\bullet$ for the partial order on
$E(A) \cup (C(A) \times \{\bullet\})$, with $(c,a) \leq (c',a')$ if $c \leq c'$in $C(A)$  and $a \leq a'$ in $V\!(A)_\bullet$.
\begin{defi}A \emph{proof} of an event $e \in E(A)_\bullet$ is an \emph{ordinal sequence} of events $(e_\alpha)_{\alpha \leq \kappa}$ in $E(A)_\bullet$ such that $e_\kappa \leq e$ and for $\alpha \leq \kappa$, where $e_\alpha = (c_\alpha,a_\alpha)$:
\begin{itemize}
\item If $\alpha = 0$ then  $* \vdash c_\alpha$ ($c_0$ is initial), 
\item If $\alpha = \beta+1$ then $e_\beta \vdash_A c_{\alpha}$,
\item If $\alpha = \bigvee_{\beta < \alpha} \beta$, then  $c_{\alpha} = \bigvee_{\beta < \alpha}c_\beta$.
\end{itemize}
For $x \subseteq E(A)_\bullet$, we write $x \vvdash e$ if there is a proof   of $e$ in $x$. If this has length $\kappa$, we write $x \vdash^\kappa e$. We write $x \vvdash c$ if $x \cup \{(c,\bullet)\} \vvdash (c,\bullet)$ (resp. $x \vdash^\kappa c$ if $x \cup \{(c,\bullet)\} \vdash^\kappa (c,\bullet)$.
\end{defi}
Note that:
\begin{itemize}
\item Every event in a  proof in $A$ except (perhaps) the last one is in $E(A)$. 
\item For any proof $(c_\alpha,a_\alpha)_{\alpha \leq \kappa}$, if $\beta < \alpha \leq \kappa$ then $c_\beta < c_\alpha$. 
\end{itemize}

\begin{defi}
A \emph{state} of an ocds $A$ is  a set of events  $x \subseteq  E(A)_\bullet$ which  satisfies:
\begin{description}
\item [Upwards Closure]If $e \in x$ and $e \leq e'$ then $e' \in x$.
\item [Safety] If $e \in x$ then there exists $e' \in x$ such that $e' \leq e$ and  $x \vvdash e'$.  
\end{description}
\end{defi}
We write $D(A)$ for the set of states of $A$. Some further conditions on states will be useful. 
\begin{defi}
A state $x\in D(A)$ satisfies:
\begin{itemize}
\item  \emph{totality} if $x \subseteq E(A)$ (i.e. no cell is filled with $\bullet$ in $x$). 
\item \emph{finite safety} if $e \in x$ implies $x \vdash^i e$ for some $i < \omega$ --- i.e. every event in $x$ has a finite proof in $x$. 
%\item \emph{infinite livelock}
\item \emph{finite-branching} if $(c,\bullet) \not \in x$ implies $\{v \in V\!(A)\ |\ (c,v) \in x\}$ is finite, and $x$ satisfies the following \emph{infinite livelock} condition:\\ 
 if  $x \vdash^\lambda c$ and $\lambda \geq \omega$ implies $(c,\bullet) \in x$ --- i.e. every cell which has an infinite proof in $x$  is filled with $\bullet$ in $x$.
% < \infty$. 
\end{itemize} 
\end{defi}
Note that any total state which satisfies infinite livelock also satisfies finite safety. 
\begin{exa}
For any set $X$, the ocds  $\widehat{X} = (\{c\},X,\{(*,c)\},  \{c\} \times X)$, has a single initial cell $c$, which may be filled with any value in $X$ (ordered discretely). Clearly, every state of $\widehat{X}$ trivially satisfies finite safety and infinite livelock  
\begin{itemize}
\item The total states of $\widehat{X}$ are in one-to-one correspondence with the subsets of $X$, 
\item The total and finite-branching states are in one-to-one correspondence with the \emph{finite} subsets of $X$. 
\item There is a single non-total (and finite-branching) state containing $(c,\bullet)$ and thus every event of $\widehat{X}$.
\end{itemize}
\end{exa}

A second key example is the following ``universal'' ocds, which will play an important role in our full abstraction result.
\begin{defi}
Let   $\U$ be the ocds with a countable set of cells $\{c.i\ |\ i \in \omega\}$, all of which are initial and may be filled by a single value $\star$. 
\end{defi}
Notwithstanding the simplicity of $\U$, for any countable ocds $A$, each  state $x \in A$ may be uniquely  represented as a state of $\U$: let $\{e_i \ |\ i \in \omega\}$ be an enumeration of $E(A)_\bullet$ and define $\inj(x) \in D(\U)$:  \\
$(c.{i},\star) \in  \inj(x)$ iff  $e_i \in x$,\\ 
$(c.{i},\bullet) \in \inj(x)$ iff $e_i \in x$, where $e_i = (c,\bullet)$ for some $c$.

The following construction illustrates the role of the partial order on the cells of an ocds. There is a subclass of ocds for which this order is given by the transitive closure of the relation: $c \ll c'$ if there exists $v$ such that $(c,v) \vdash c'$. In fact, this forms a sub-CCC of our category of ocds. However, it is not closed under the following \emph{lifting} operation, which adds an unfillable cell which can therefore not enable any other cells.
\begin{defi}[Stable Lifting] For  any ocds $A$, let $A_\up = (C(A)_\bot,
  V\!(A), \vdash_A \cup \{(*,c_\bot)\}, \allowbreak E(A))$, where $C(A)_\up$ is the lifting of $A$ by adding a new least element $c_\bot$. 
\end{defi}
 % Lifting adds an initial  cell to $A$ which is not filled by any value.   
Any state of $A_\up$ which contains $(c_\bot,\bullet)$ must contain all events of $E(A)$ --- i.e.  the states of $A_\up$ are the states of $A$, together with one new  state containing all possible events.

For a state $x \in D(A)$, we identify the following sets of cells  (subsets of  $C(A)$):
\begin{itemize}
\item $F(x) = \{c \in C(A)\ |\ \exists a\in V\!(A)_\bullet.(c,a) \in x\}$ --- the set of \emph{filled} cells of $x$.
\item $En(x) = \{c \in C(A) \ |\ x \vvdash c\}$ --- the set of \emph{enabled} cells of $x$.   
\item $A(x)  = En(x) - F(x)$ --- the set of \emph{accessible}   cells of $x$.
 \end{itemize} 
If $c \in En(x)$, and $(c,a) \in E(A)_\bullet$,  we write $x + (c,a)$ for the state $x \cup \{e \in E(A)_\bullet\ |\ (c,a) \leq e\}$. We will also  write $x + (c,V)$ for $\bigcup_{v \in V}(x + (c,v))$.

Say that a state $x$ is \emph{complete} when every cell with a finite proof in $x$ is filled.  
\begin{exa}Consider the ocds of \emph{lazy} natural numbers: the set of cells is $\{c_i \ |\ i \leq \omega\}$, ordered by $i$.  For each $i < \omega$, $c_i$ can be filled with either of the values $0$ or $\succ$. $c_0$ is initial and $c_{i+1}$ is enabled by $(c_i,\succ)$.   

Proofs are ascending chains of events $(c_0,\succ),(c_1,\succ),\ldots$, and thus: 
\begin{itemize}
\item The total, complete and finitely safe states are in one-to-one correspondence with the non-empty subsets of $\Na$.
\item The total and complete finitely branching states are in one-to-one  correspondence with the non-empty finite subsets of $\Na$ (since any state in which infinitely many cells are filled must contain a proof of $(c_\omega,\bullet)$ by the infinite livelock condition).  
\end{itemize}
The states which are either non-total or incomplete correspond to a subset of $\Na$  bounded above by $i \leq \omega$, together with a proof of $c_i$, which is either filled with $\bullet$ or (for $i < \omega$) unfilled. (Filling $c_\omega$ with $\bullet$ corresponds to presence of the recursively defined element $n := \suc(n)$.) 
\end{exa}

\subsection{Ordered  Concrete Data Structures as Biorders}
For each ocds, we define a biorder in which the set of elements is the set $D(A)$ of states of $A$. The extensional order is set-inclusion: since any union of states satisfies the safety and upwards-closure conditions, $(D(A),\subseteq)$ is a complete (join) lattice, with least element the empty state and greatest element $\bigcup D(A)$.
\begin{defi}The stable order on states is defined as follows:
\begin{center}
$x \leqs y$ if  $y \subseteq x$ and for all   $(c,v) \in x$, either $(c,v) \in y$ or $(c,\bullet) \in x$. 
\end{center}
\end{defi}
 $(D(A),\subseteq,\leqs)$ is a biorder --- if $X$ is a non-empty set of states bounded above by $y$, then $\bigcup X$ is their greatest lower bound in the stable order: if $(c,v) \in \bigcup X$ then $(c,v) \in x$ for some $x \in X$ and either $(c,\bullet) \in x \subseteq \bigcup X$ or else $(c,v) \in y$ and so for all $x' \in X$, $(c,v) \in x'$.
If $z \leqs x$ for all $x \in X$ then $\bigcup X \subseteq z$ and if  $(c,v) \in z$  either $\forall x.(c,v) \in x \in \subseteq \bigcup X$ or $(c,\bullet) \in z$.     $(D(A),\subseteq,\leqs)$ is pointed --- its stably least element is $\bigcup D(A)$.

It is easy to see that $D(A)$ is extensionally complete, but what of stable completeness? In fact, there are  pathological examples  of ocds in which this property does not hold, but we will show that the ocds for which $D(A)$ \emph{is} stably complete  form a Cartesian closed category. Note that in any case, if $\Delta$ is a non-empty set of states with a stable upper bound then its stable l.u.b. is $\bigcap \Delta$. 

Evidently, the sub-biorder  $D^*(A)$ consisting of  states which are finitely safe is also an (extensionally complete) biorder, as this property is preserved by arbitrary unions of states.   The sub-biorder $D_*(A)$ consisting of finite branching states is a biorder, although not, in general extensionally complete, since the infinite  union of finite-branching states may not be finite-branching. However, it is stably complete.
\begin{lem}If $X\subseteq D_*(A)$ is stably bounded above by $y \in D_*(A)$, then $\bigcup X \in D_*(A)$.
\end{lem}
\begin{proof} Suppose there is a cell $c$ and infinite set of values $V$  such that for all $v \in V$, $(c,v) \in \bigcup X$. If every such event occurs in $y$, then $(c,\bullet) \in y \subseteq \bigcup X$. Otherwise there exists $x \in X$ with $(c,v) \in x$ and $(c,v) \not \in Y$ and so $(c,\bullet) \in  x \subseteq \bigcup X$.
Suppose there is an event $(c,\bullet)$ which has an infinite proof in $\bigcup X \cup \{(c,\bullet)\}$. If this proof is in $y$ then $(c,\bullet) \in y \subseteq X$. Otherwise there is an event $(c',v)$ in the proof such that $(c,v) \not \in y$, and hence $x \in X$ such that $(c', \bullet) \in x \subseteq \bigcup X$ such that $(c',\bullet) \in x$ and so $(c,\bullet) \in \bigcup X$ by upwards closure.
\end{proof} 

We shall  use the following  alternative characterizations of the stable order and stable boundedness.
\begin{defi}Given a state $x$ and  $C \subseteq En(x)$, let $x_C = \bigcup  \{x + (c,\bullet) \ |\ c \in C\}$. 
\end{defi}
\begin{lem}$y \leqs x$ iff $y = x_C$ for some  $C \subseteq En(x)$.
\end{lem} 
\begin{proof}Evidently, $x_C \leqs x$, so it suffices to show that every element $y \leqs x$ has this form. Let $C = \{c \in En(x)\ |\ (c,\bullet) \in y\}$, so that $x_C \subseteq y$.   We claim that  $y \subseteq x_C$: suppose $(c,a) \in y$, but $(c,a) \not \in x$. Let $\langle (c_\alpha,v_\alpha)\ |\ \alpha  <\kappa \rangle$ be a proof of $(c,a)$ in $y$. Then either $(c,a) \in x$ or there exists a least value $\alpha$ such that $(c_\alpha,v_\alpha) \not \in x$. So $c_\alpha \in E(x)$, and $(c,\bullet) \in y$ by stability and so $c_\alpha \in C$. Since $c_\alpha \sqleq c$,  $(c,a) \in x_C$ as required. 
\end{proof}

\begin{lem}A non-empty set $X$ of states is stably bounded above if (and only if):
\begin{center}$(\dagger)$\ \ \ \ for all $(c,v) \in x \in X$, either $(c,\bullet)  \in x$ or $(c,v) \in \bigcap X$.
\end{center}
\end{lem}
\begin{proof}
From left-to-right, this follows from the fact that $X$ is bounded above by $\bigcap X$. From right-to-left, we need to show that if $X$ satisfies $(\dagger)$ then  $\bigcap X$ satisfies the safety condition, since it is then a stable upper bound. Suppose $(c,v) \in  \bigcap X$ --- choose $x \in X$  and let  $P$ be a shortest proof of $(c,a)$ in $x \in X$ (i.e. there is no shorter proof). Then $P \subseteq \bigcap X$, since for every preceding element  $(c',v')$ in $P$ we have $(c',\bullet) \not \in x$ (or else a shorter proof would exist) and so $(c',v') \in \bigcap X$.   
\end{proof}
Returning to our examples, we may observe that:
\begin{itemize} 
\item For any set $X$, $D(\widehat{X}) \cong \xp(X)_\up$ and $D_*(\widehat{X}) \cong\xp_{\mathit fin}(X)_\up$. In particular, $D(\widehat{\varnothing})$ is the two-element space $\Sigma$.
\item $D(\U) \cong \[U\]$, where $U = \nt \rightarrow \o \rightarrow \o$ --- it is the countable product of copies of $\[o \rightarrow \o\] \cong \widehat{1}$. Moreover, for any ocds $A$, the map sending $x \in D^*(A)$ to $\inj(x)$ is a continuously stable function, with a left inverse --- $\proj(y) = \bigcap\{x \in D^*(A) \ |\ \inj(x) \leqs y\}$ --- i.e.   $\U$ is a \emph{universal object} among the countable ocds and continuously stable functions.    
\item For any ocds $A$, $D(A_\up) \cong D(A)_\up$ (and $D^*(A_\up) \cong D^*(A)_\up$, $D_*(A_\up) \cong D_*(A)_\up$) --- lifting adds a single element which is stably least/extensionally greatest.  
\end{itemize}

\section{Sequential Algorithms on Ordered Concrete Data Structures}
\label{sec:s5}
By the results in the previous section, we may define the following categories:
\begin{itemize}
\item  $\ocd$ --- objects are ordered concrete data structures, morphisms from $A$ to $B$ are monotone stable functions from $D(A)$ to $D(B)$.
\item $\ocde$ --- objects are  ocds and morphisms from $A$ to $B$ are  continuously stable functions from $D^*(A)$ to $D^*(B)$.
\item $\ocds$ --- objects are  ocds and  morphisms  from $A$ to $B$  are  stably continuous functions from $D_*(A)$ to $D_*(B)$.
\end{itemize}
 Each category has  products, given by the disjoint union of ocds: $\Pi_{i \in I}(C_i,V_i,E_i,\vdash_i) =\\ (\coprod_{i \in I} C_i, \coprod_{i \in I} V_i,\bigcup_{i \in I}\{(c.i,v.i)\ |\ (c,v) \in E_i\}, \bigcup_{i \in I}\{((c.i,v.i),d.i)\ |\ ((c,v),d) \in E_i\})$.\\
The full, identity-on-morphisms functor  $D: \ocd \rightarrow \B$ sending each  concrete data structure to its set of states (which restricts to functors $D^*:\ocde \rightarrow \EB$ and $D_*:\ocds \rightarrow \SB$) preserves products, and
so to establish Cartesian closure  in each case, it remains to define an internal hom ocds for each pair of objects and show that its biorder of states is isomorphic to the internal hom (function space) in the corresponding category of biorders.
This is a key result, since it establishes that every stable and monotone function between the biorder of states of an ocds is computed by a unique state of this exponential ocds or \emph{sequential algorithm}.

Focussing first on the general case of  monotone stable functions,  for each pair of ocds $A,B$ we define an ordered concrete data structure $A \Rightarrow B$ (cf. the analogous definition of  unordered concrete data structure \cite{BeCu})  and show that  $D(A \Rightarrow B) \cong [D(A),D(B)]_{\B}$ in  $\B$.  
\begin{description}
\item [Cells] A cell of $A \Rightarrow B$ is given by a pair of a total state of $A$ and a cell of $B$:\\
$C(A \Rightarrow B) = (D(A) \cap {\cal{P}}(E(A)))  \times C(B)$ --- with $(x,c) \leq (x',c')$ if $x \subseteq x'$ and $c \leq c'$.
\item [Values] A \emph{value} of $A \Rightarrow B$ is either a cell from $A$ or a value from $B$ --- the order being determined pointwise from that of $V\!(B)$ and the reverse of $C(A)$: \\ $V\!(A \Rightarrow B) = C(A)^{c} \uplus V\!(B)$ 
\item [Events] A cell $(x,c)$ of $A \Rightarrow B$ may be \emph{filled} with either a cell accessible from $x$ in $A$ or a value in $B$ which can fill $c$:\\
$E(A \Rightarrow B)  = \{((x,c),c'.1)\ |\ (x,c) \in C(A\Rightarrow B) \wedge c' \in A(x)\} \cup \{((x,c),v.2)\ |\ (x,c) \in C(A\Rightarrow B) \wedge (c,v) \in E(B)\}$

\item [Enabling] An event $((x,c),c'.1)$ enables a cell $(x',c)$ if $x'$  may be obtained from $x$ by filling $c'$, and an event $((x,c),v.2)$ enables a cell $(x,c')$ if $(c,v)$ enables $c'$  in $B$: 
$\vdash_{A \Rightarrow B} =$\\$ \{(((x,c),c'.1),(x',c)) \in E(A \Rightarrow B) \times C(A \Rightarrow B)\ |\ \exists V \subseteq  V\!(A).x' = x + (c',V)\} \cup \{(((x,c),v.2),(x,c')) \in E(A \Rightarrow B) \times C(A \Rightarrow B)\ |\ (c,v) \vdash_B  c'\} $    
\end{description}
A \emph{sequential algorithm} from $A$ to $B$ is a state of $A \Rightarrow B$. In the following, we will assume that the sets of cells of $A$ and values of $B$ are disjoint, and so omit explicit tagging. 

\subsection{Stable Functions from Sequential Algorithms} 
% We need to establish that $D(A \Rightarrow B)$ and $[D(A),D(B)]$ are isomorphic in $\B$. 
Every sequential algorithm $\sigma \in D(A \Rightarrow B)$ computes  a monotone stable function $\fun(\sigma)$ from $D(A)$ to $D(B)$. Given a  state $x \in D(A)$, define:
$\fun(\sigma)(x) =$ $$ \{(c,a)\in E(B)_\bullet\ |\ \exists x' \subseteq x.((x',c),a) \in \sigma \vee \exists c'.(c',\bullet) \in x \wedge ((x',c),c') \in \sigma\}$$  
\begin{lem}For any $x\in D(A)$, $\fun(\sigma)(x)$ is a state in $D(B)$.
\end{lem}
\begin{proof} $\fun(\sigma)(x)$ satisfies:
\begin{description}
\item [Upwards closure] Suppose $(c',a') \geq (c,a) \in \fun(\sigma)(x)$. If there  exists $x' \subseteq x$ with $((x',c),a) \in \sigma$ then  $((x',c'),a') \geq ((x',c),a)$ and so $((x',c'),a') \in \sigma$ and $(c',a')\in \fun(\sigma)(x)$.\\
If there exists $(c'',\bullet) \in x$ such that $((x',c),c'') \in \sigma$ then $((x,c'),c'') \geq   ((x',c),c'') $ so $ ((x',c'),c'') \in \sigma$ and hence  $(c',a')\in \fun(\sigma)(x)$.
\item [Safety]Suppose $(c,a) \in \fun(\sigma)(x)$. Then then there exists  $((y,c),b) \in  \sigma$ with $y \subseteq x$, and a proof of $(y,c)$ in  $\sigma$ which therefore restricts to a proof of $c$ in $f(y)$. \qedhere
\end{description}
\end{proof}
Say that a function $f:D(A) \rightarrow D(B)$ is computed by the sequential algorithm $\sigma$ if $f(x) = \fun(\sigma)(x)$ for all $x \in D(A)$. We now show that any such function is  monotone  stable. % (and later, that every monotone stable function is computed by a unique sequential algorithm).
\begin{lem}\label{l1}If $\up X$ and $(c,a) \in \fun(\sigma)(\bigcup X)$ then either $(c,a) \in \fun(\sigma)(x)$ for all $x \in X$ or $(c,\bullet) \in \fun(\sigma)(x')$ for some $x' \in X$.
\end{lem}
\begin{proof}
If  $(c,a) \in \fun(\sigma)(\bigcup X)$ then  there exists an event $e \in \sigma$ such that either  $e = ((w,c),a)$, where $w \subseteq \bigcup X$  or $e = ((w,c),c')$, where  $w + (c',\bullet) \subseteq \bigcup X$. If $w \subseteq x$ for every $x \in X$ then in the first case $(c,a) \in \bigcup\fun(\sigma)(x)$ for every $x \in X$, and in the second case there exists $x' \in X$ with $w + (c',\bullet) \subseteq x'$ and so $(c,\bullet) \in \fun(\sigma)(x')$.  

So suppose $w \not \subseteq x$ for some $x \in X$. Fixing a proof of $e$ in $\sigma$,  let $e' = ((w',c'),a')$ be the first element in this proof such that  $w' \not \subseteq x$ for some $x \in X$. Then there is an immediately preceding event $((w'',c'),c'')$ such that $w' = w'' +(c'',V)$ for some set of values $V$, including a value $u$ such that $(c'',u) \not \in x$.  Because $w' \subseteq \bigcup X$, there exists $x' \in X$ with $(c'',u) \in x'$. Since $x\up x'$, therefore   $(c'',\bullet) \in x'$,  and hence $(c',\bullet) \in \fun(\sigma)(y)$. Since $c'\leq c$, we have  $(c,\bullet) \in \fun(\sigma)(x')$ as required.
\end{proof}

\begin{prop}\label{wdef} For any sequential algorithm $\sigma$, $\fun(\sigma)$ is a monotone stable function. 
\end{prop}
\begin{proof}
Evidently, if $x \subseteq y$ then $\fun(\sigma)(x) \subseteq \fun(\sigma)(y)$. 
%If $x \leqs y$, then $f(x) \leqs f(y)$: suppose $(c,v) \in f(x) = f(x \cup y)$ then by Lemma~\ref{l1}, $(c,v) \in f(y)$ or $(c,\bullet) \in f(y)$.  
If $\up X$, then $\fun(\sigma)(\bigcup X) = \bigcup_{x \in X} \fun(\sigma)(x)$: if $(c,a) \in  \fun(\sigma)(\bigcup X)$ then by Lemma~\ref{l1}, either $(c,a) \in \fun(\sigma)(x)$ for all $x \in X$ or $(c,\bullet) \in \bigcup_{x \in X} \fun(\sigma)(x)$. 
\end{proof}
\subsection{Examples}
We now describe the sequential algorithms which compute some of the monotone stable  functions described in the examples of Sections~\ref{sec:s2} and~\ref{sec:s3}.
\begin{exa}For any set $X$, consider the ocds $\widehat{\varnothing}^X \Rightarrow \widehat{\varnothing}$ --- i.e. $\widehat{\varnothing}^X$ is the $X$-indexed product of copies of $\widehat{\varnothing}$, which has (initial) cells $\{c_x\ |\ x \in X\}$ and no values,  so $\widehat{\varnothing}^X \Rightarrow \widehat{\varnothing}$ has a single initial cell $(\{\_\},c)$, which may be filled by any of the cells $c_x$ for $x \in X$. In other words, it is the same ocds as $\widehat{X}$, up to renaming of cells and values, and so $D(\widehat{\varnothing}^X \Rightarrow \widehat{\varnothing}) \cong D(\widehat{X}) \cong \xp(X)_\up \cong [\widehat{\varnothing}^X,\widehat{\varnothing}]_{\B}$ by Proposition \ref{pd}, and $D_*(\widehat{\varnothing}^X \Rightarrow \widehat{\varnothing}) \cong D_*(\widehat{X})   \cong \xp_{\mathit{fin}}(X)_\up \cong [\widehat{\varnothing}^X,\widehat{\varnothing}]_{SB}$. 
\end{exa}
By considering the algorithms computing the functions of Example \ref{nfa}, we may see that unlike a  non-deterministic strategy in e.g. the games model of \cite{HM},  a non-deterministic sequential algorithm may have different decompositions into a non-deterministic choice between two (complete and total) algorithms. 
\begin{exa}
Consider the OCDS $\widehat{\mathbb B} \times \widehat{\mathbb B} \Rightarrow \widehat{\mathbb B}$.
\begin{itemize}
\item Cells are pairs $(x,c)$, where $x$ is a total state of $\widehat{\mathbb B} \times \widehat{\mathbb B}$ (corresponding to a pair of subsets of $\B$).
\item Values are either a Boolean (filling $(x,c)$ for any $x$) or a cell ($c_0$ or $c_1$) of
$\widehat{\mathbb B} \times \widehat{\mathbb B}$, filling $(x,c)$ provided it is not filled in $x$.   
\end{itemize}
So a sequential algorithm  in $\widehat{\mathbb B} \times \widehat{\mathbb B} \Rightarrow \widehat{\mathbb B}$ may investigate (both, either or neither) of its arguments in any order, and return a Boolean value. The functions denoted by the terms $t_{i,j}$  of Example \ref{nfa} are computed by sequential algorithms $\sigma_{i,j}$ as follows:
$$\xymatrix@R=8pt{ \widehat{\mathbb B} \times \widehat{\mathbb B} & \Rightarrow & \widehat{\mathbb B} \\
& (\{\_\},c)\\
c_i \\
& (\{(c_i,v_i)\},c)\\
c_{i^*} \\
& (\{(c_0,v_0),(c_1,v_1)\},c)\\
& & v_j}$$
Observe that the cell $(\{(c_0,v_0),(c_1,v_1)\},c)$ does not depend on $i$ or $j$, and thus $\sigma_{0,0} \cup \sigma_{1,1} = \sigma_{0,1} \cup \sigma_{1,0}$. 
\end{exa}

In fact, we can present sequential algorithms from $A$ into $\widehat{\varnothing}$ (which might  be considered as continuations on $A$) more simply by ignoring the unfillable cell of $\widehat{\varnothing}$ --- i.e. cells correspond to states of $A$ ordered by inclusion,  values to cells of $A$, and an event to a pair $(x,c)$ of  a state of $A$ and a cell which is accessible from it, which enables the states (new cells) of the form $x + (c,V)$.  The sequential algorithms computing the functions of Example \ref{bta} may be presented in this way: 
\begin{exa}
Consider the concrete data structure $(\widehat{\mathbb B} \Rightarrow \widehat{\varnothing}) \Rightarrow \widehat{\varnothing}$. Ignoring the unfillable cell of $\widehat{\varnothing}$, its initial cell is the empty state, which can be  filled (only) by the initial cell of $\widehat{\mathbb B} \Rightarrow \widehat{\varnothing}$ (also the empty state). $(\varnothing,\varnothing)$ enables the cell (state of $ \widehat{\mathbb B} \Rightarrow \widehat{\varnothing}$) $\{(\varnothing,c)\}$, where $c$ is the (unique) initial cell of $\widehat{\mathbb B}$, which may be filled with the values (non-empty states of $\widehat{\mathbb B}$) $\{(c,\tru)\}$, $\{(c,\ff)\}$ and $\{(c,\tru), (c,\ff)\}$. There is a non-trivial order on these values: $\{(c,\tru)\}, \{(c,\ff)\} \geq \{(c,\tru), (c,\ff)\}$, so by upwards closure any cell which is filled with the last must also be filled with the first two. This distinguishes the sequential algorithms computing   $\lambda f.f\spc (\tru + \ff)$ and $ \lambda f.(f\spc \tru) + (f\spc \ff)$: in the first case 
 $\{(\varnothing,c)\}$ is filled with $\{(c,\tru), (c,\ff)\}$,  in the second case it is not. (See Figure \ref{lfa} for the extensional hierarchy of sequential algorithms at this type.)
\end{exa}
\begin{figure}
\begin{center}
$ \varnothing \  (\bot_E)$ \\
$\{(\varnothing,\varnothing)\} \ (\lambda f.f\spc \bot_E)$\\
 $\{(\varnothing,\varnothing),((\varnothing,c),\{(c,\tru)\})\} \ (\lambda f. f\spc \tru) \ \ \ \ \ \   \ \ \ \ \  \ \ \ \ \ \  \{(\varnothing,\varnothing),((\varnothing,c),\{(c,\ff)\})\}\  ( \lambda f.f \spc \ff)$\\ 
  $\{(\varnothing,\varnothing),((\varnothing,c),\{(c,\tru)\}),((\varnothing,c),\{(c,\ff)\})\} \  (\lambda f.(f\spc \tru) + (f\spc \ff))$\\
  $\lfloor \{(\varnothing,\varnothing),((\varnothing,c),\{(c,\tru),(c,\ff)\})\}\rfloor \ (\lambda f.f\spc (\tru + \ff))$\\
  $\lfloor \{(\varnothing,\varnothing),((\varnothing,c),\bullet)\}\rfloor\   (\lambda f.f\spc \top_E)$\\
$ \lfloor \{(\varnothing,\bullet)\} \rfloor \ (\top_E)$\\
\caption{Sequential Algorithms on $(\widehat{\mathbb B} \Rightarrow \widehat{\varnothing}) \Rightarrow \widehat{\varnothing}$ (ordered extensionally)}\label{lfa}
\end{center}
\end{figure} 
%\end{exa}
The above examples relate to finitary (or, at least, finitely safe) sequential algorithms. Recasting Example \ref{ncs} furnishes an instance of an event with an infinite proof which is required to separate the sequential algorithms computing the terms $s$ and $t$ defined there.  
\begin{exa}
Consider the ocds $\U \Rightarrow \widehat{\varnothing}$: again, ignoring the unfillable cell on the right this has as cells total states of $\U$ (which correspond to subsets of $\Na$), and as values, cells of $\U$ (which correspond to elements of $\Na$) --- an event is thus a pair $(X,x)$ such that $x \not \in X$, with $(X,x) \vdash Y$ iff $Y = X \cup \{x\}$. A  sequential algorithm on $\U \Rightarrow \widehat{\varnothing}$ thus corresponds to a set of finite non-repeating sequences over $\Na$, together with a set of divergences  ---  finite and infinite sequences over $\Na$ ending with a cell filled with $\bullet$ --- such that any sequence which extends a divergence  appears in both sets. Compare with Harner and McCusker's game semantics of non-determinism \cite{HM}.      

The sequential algorithms on $\U \Rightarrow \widehat{\varnothing}$ which compute the functions denoted by $s$ and $t$ defined in Example \ref{ncs} contain the same  finite sequences of events (corresponding to finite prefixes of $0123\ldots$). However, the algorithm computing $s$ contains a divergence corresponding to the infinite sequence  $0123\ldots$, whereas the algorithm computing  $t$ does not. To be more precise, for each  $j \leq \omega$, let $P_{j}$ be be the state  $\{(c.i,\star)\ |\ i < j\}$ in $D(\U)$, and for each $k \leq \omega$ let $Q_{k}$ be the (upwards closure of) $\{((P_j,c),c_{j})\ |\ j < k\}$ in $D(\U \Rightarrow \widehat{\varnothing})$.  Then $\[s\]$ is computed by $Q_\omega  \cup \{((P_\omega,c),\bullet)\}$ and $\[t\]$ is computed by  $Q_{\omega}$ --- i.e. they are differentiated only by the event $((P_\omega,c),\bullet)$, which is the conclusion of the infinite  proof $((P_j,c),c_{j})_{j < \omega}$ in $Q_\omega$.

 One might wonder whether the only infinite proofs which are needed to distinguish terms of $\Lambda(N)^+$  are the ones that lead to a divergence in this way. However, at the next  order of types (third order) we may give examples of terms denoting \emph{total} sequential algorithms which are distinguished only by events with infinite proofs. Define $p:(U \rightarrow \o) \rightarrow \o  \triangleq \lambda g.g\spc \lambda u.\lambda x.x$, and $r:(U \rightarrow \o) \rightarrow \o  \triangleq \lambda u.\lambda x.?_N\spc \lambda v.({\mathtt{leq}}(u,v)\spc x)\spc \mho$, so that  $\[p\]$ is computed by applying its argument to the infinite state $P_\omega \in D(\U)$ and  $\[r\]$ by non-deterministically applying its argument to some finite  state $P_j$ in $\U$. Writing $\(p\)$ and $\(r\)$ for these sequential algorithms:
\begin{itemize}
\item every event in $\(p\)$ with a \emph{finite proof} (i.e. the finite states $P_j$ for $j < \omega$)  also occurs in $\(r\)$.
\item  $p$ and $r$  are distinguished by application to $t$ ---
 $p \spc t \not \Downarrow_\must$ and $r\spc t \Downarrow_\must$, since $\(p\)$ contains the event  $((Q_\omega,c),P_\omega)$ --- which has the proof $((Q_k,c),P_k)_{k \leq \omega}$ in $\(p\)$ --- but  $\(r\)$ does not. 
\end{itemize}
We may extend these examples to give terms which are distinguished only  by events with proofs of length  $\omega + l$, for any $l < \omega$.
Defining $U^0 \triangleq U$ and $U^{l+1} \triangleq (U \rightarrow \o) \rightarrow \o$, we give terms $p^l:U^l$ and  $t^l,s^l:U^l \rightarrow \o$ by $p^0 \triangleq \lambda u.\lambda x.x$ and $p^{l+1} \triangleq \lambda f.f\spc p^l$,   
$t^0 \triangleq t$ and $t^{l+1} \triangleq \lambda f.f\spc t^l$ and   $s^0 \triangleq s$ and $s^{k+1} \triangleq \lambda f.f\spc s^k$. Then $t^l\spc p^l \not\Downarrow_\must$ for each $l$, but $s^l\spc p^l \Downarrow_\must$.     

Denotationally, for each $l\in \omega$, $\(t^l\) = \(s^l\) \cup \{((P^l_{\omega +l},c),\bullet)\}$, where  $P^l_j \in D(\U^l \Rightarrow \widehat{\varnothing})$ and $Q^l_j \in D\((\U^l \Rightarrow \widehat{\varnothing}) \Rightarrow \widehat{\varnothing} \)$   are defined inductively for $1 \leq j \leq \omega + l$ by $P^0_j = P_j$ and $Q^0_j = Q_j$, and   $P^{l+1}_j = \{((Q_i^l,c),P^l_{i})\ |\ i < j\}$ and $Q^{l+1}_j = \{((P_i^{l+1},c),Q^l_i)\ |\ i < j\}$. 

Observe that the proof of  $(P_{\omega +l},c)$ in $\[t^l\]$ is the $\omega + l$ sequence of events $((P_i^{l+1},c),Q^l_i)_{j < \omega +l}$.
In the simply-typed language $\Lambda(N)^+$, it appears that this is as far as it is necessary to go --- we conjecture that any inequivalent terms at a type of order at most $2l$ may be distinguished by an event with a proof no longer than $\omega + l$. 
\end{exa}

\section{Stable Functions and Sequentiality}
\label{sec:s6}
We now return to the task of showing that the functions between the biorders of states of ocds which are computed by sequential algorithms are precisely the monotone and stable ones. 
Recall that concrete data structures were originally  introduced by Kahn and Plotkin in order to give a description of sequentiality for  higher-order deterministic  functionals  \cite{KP}. Essentially, in this  setting a function between the states of concrete data structures $A$ and $B$ is Kahn-Plotkin sequential if any argument (state of $A$) $x$,   and cell $c$ of $B$ which is filled in  $f(y)$ for some $y$ which extends $x$,   can be associated with a cell, accessible from $x$, which must be filled in any state $z$ (which extends $x$) such that $c$ is filled in $f(z)$.  However, in this original setting, divergence is represented \emph{implicitly}, by not filling an enabled cell (rather than as an explicit divergence  by filling a cell with $\bullet$), and inclusion of states corresponds to the \emph{stable} order. Thus we  translate this original definition of Kahn-Plotkin sequentiality to the current setting by (essentially) replacing the role of ``accessible cell'' with that of ``cell filled with $\bullet$'', and ``filled cell'' with ``enabled cell not filled with $\bullet$''.   We define a partial order ($\preceq$) on total states (which plays the role of the stable order in the original definition of Kahn-Plotkin sequentiality): 
\begin{center} $x \preceq y$ if $x \subseteq y$ and if $c \in F(x)$ then $(c,v) \in y$ implies $(c,v) \in x$.
\end{center}
\begin{defi}A function $f:D(A) \rightarrow D(B)$ is \emph{explicitly sequential}  if whenever $x,y$ are total states such that $x \preceq  y$  then for any event $(c,v) \in f(y)$, either $(c,v) \in f(x)$, or there exists $c' \in A(x) \cap F(y)$ such that if  $\up\{ x,z\}$ and  $(c', \bullet) \in z$  then $(c,\bullet) \in f(z)$.  
\end{defi}
We will now show that all monotone stable functions are explicitly sequential.
\begin{lem}If $x \preceq y$ then $y \subseteq x_{A(x) \cap F(y)}$.
\end{lem}
\begin{proof}
Suppose $(c,v) \in y$ but $(c,v) \not \in x$. Let $P$ be a proof of $c$ in $y$, and $(c',v')$ the least element of $P$ which is not in $x$. Then $c'$ is accessible (enabled but not filled) in $x$, as $x \preceq y$ --- i.e. $c' \in A(x)\cap F(y)$, so $(c,v) \geq (c',\bullet) \in x_{A(x)\cap F(y)}$. 
\end{proof}
\begin{prop}\label{KP}Every monotone stable function $f:D(A) \rightarrow D(B)$ is explicitly sequential.
\end{prop}
\begin{proof}
Suppose $x \preceq y$ and  $(c,v) \in f(y)$ but $(c,v) \not \in f(x)$. Since $y \subseteq x_{A(x) \cap F(y)}$, $(c,v) \in f(x_{A(x) \cap F(y)})$ and as $f(x_{A(x) \cap F(y)}) \leqs f(x)$, so $(c,\bullet) \in f(x_{A(x) \cap F(y)})$. 

By conditional multiplicativity, $f(x_{A(x) \cap F(y)}) = \bigcup_{c' \in A(x) \cap F(y)} f(x + (c',\bullet))$ and so  there is  a cell $c'  \in A(x) \cap F(y)$ such that $(c,\bullet) \in f(x + (c',\bullet))$ as required.   
\end{proof}
We establish the converse --- that every explicitly sequential function is monotone stable~--- by showing below that every explicitly sequential function from $D(A)$ to $D(B)$ is computed by a sequential algorithm.

\subsection{Sequential Algorithms from Monotone Stable Functions}
We will now  use the explicit sequentiality property to establish  that every monotone stable function  $f:D(A) \rightarrow D(B)$ is computed by a   sequential algorithm $\strat(f) \in D(A \Rightarrow B)$.
Define this   to be the set of events: 
\begin{center}
$\{((x,c),a) \in C(A \Rightarrow B) \times V\!(B)_\bullet \spc |\spc  (c,a) \in f(x)\}$\\ $\cup \{((x,c), c')  \in C(A \Rightarrow B) \times C(A)\spc |\spc (c,\bullet) \in f(x + (c',\bullet))\}$
\end{center}
\begin{lem}$\strat(f)$ is an upper set.
\end{lem}
\begin{proof}
Suppose  $((x',c'),a') \geq  ((x,c),a) \in \strat(f)$. If $a,a' \in V\!(B)_\bullet$ --- i.e. $(c,a) \in f(x)$ --- then $(c',a') \geq (c,a) \in f(x') \supseteq f(x)$, and so $((x',c'),a') \in \strat(f)$. If $a,a' \in C(A)$, so that $a' \leq a$ and hence  $x + (a,\bullet) \subseteq x' + (a',\bullet)$, then $(c,\bullet) \in  f(x' + (a',\bullet))$ and so $((x',c'),a') \in \strat(f)$ as required. 
\end{proof}

\begin{lem}\label{sf}$\strat(f)$ satisfies the safety property.
\end{lem}
\begin{proof}
We construct a proof  of each event $((x,c),a)$ in $\strat(f)$  using the explicit sequentiality property for $f$. Suppose $a \in V\!(B)_\bullet$: let  $\langle (c_\beta,v_\beta) \ |\ \beta\leq \lambda \rangle $ be a proof of $(c,a)$ in $f(x)$.

For each ordinal $\alpha$, we define:
\begin{itemize}
\item A state $x_\alpha \preceq x$.
\item An ordinal $\kappa(\alpha) \leq \alpha$ 
\item An event $e_\alpha \in E(A \Rightarrow B)$ such that $e_\alpha = ((x_\alpha,c_{\kappa(\alpha)}),v_{\kappa(\alpha)})$ or  $e_\alpha = ((x_\alpha,c_{\kappa(\alpha)}),c')$ for some $c' \in F(x)$.
\end{itemize}
such that if $\kappa(\alpha) < \lambda$, then $\langle e_\gamma\ |\ \gamma \leq \alpha\rangle$ is a proof of $e_\alpha$.
\begin{itemize}
\item Let $x_0 = \{\}$ and $\kappa(0) = 0$, 
\item If $\alpha = \bigcup\{\beta < \alpha\}$ then $x_\alpha = \bigcup_{\beta < \alpha}x_\beta$ and $\kappa(\alpha) = \bigcup\{\kappa(\beta)\ |\ \beta<\alpha\}$,  
\item For all $\alpha$, if $(c_{\kappa(\alpha)},v_{\kappa(\alpha)}) \in f(x_{\alpha})$ then let $e_{\alpha} = ((x_\alpha,c_{\kappa(\alpha)}),v_{\kappa(\alpha)})$  and $x_{\alpha+1} = x_\alpha$ and $\kappa(\alpha+1)  = {\mathtt{min}}\{\lambda,\kappa(\alpha) +1\}$.\\
Otherwise,  by the explicit sequentiality of $f$ (Proposition \ref{KP}) there is  a cell $c'  \in A(x_{j}) \cap F(x)$ such that $(c_{\kappa(\alpha)},\bullet) \in f(x_{\alpha} + (c',\bullet))$ and so we may set $e_\alpha =  ((x_\alpha,c_{\kappa(\alpha)}),c')$, and  $x_{\alpha+1} = \bigcup\{x_\alpha + (c',v)\ |\ (c',v) \in x\}$ and $\kappa(\alpha+1) = \kappa(\alpha)$. 

\end{itemize}
Since $\kappa(\alpha) \not = \lambda$  implies $e_\alpha \not = e_\beta$ for all $\beta < \alpha$, there must be some (least) $\alpha$ (no greater than the cardinality of $E(A \Rightarrow B)$) such that $\kappa(\alpha) = \lambda$ and  so $\langle e_j\ |\ \beta \leq \alpha \rangle$ is a proof of $((x,c),a)$ in $\strat(f)$. The case in which  $a$ is a cell in $C(A)$ is similar.  
 \end{proof} 
Hence we have shown that:
\begin{prop}For any monotone  stable function $f:D(A) \rightarrow D(B)$, $\strat(f)$ is a well-defined state of $A \Rightarrow B$.
\end{prop}
We now show that $\fun$ and $\strat$ are inverse --- i.e. a monotone stable function $f$ is computed by $\sigma \in D(A\Rightarrow B) $ if and only if $\sigma = \strat(f)$. Let $x^\top = \{(c,v) \in x\ |\ (c,\bullet) \not \in x\}$ (the maximal total state contained in $x$).
\begin{lem} $x^\top$ is a well-defined  state. 
\end{lem}
\begin{proof}
$x^\top$ is evidently an upper set. For safety, suppose $(c,v) \in x$: for any  event $(c',v')$ in any proof of $c$, $c' \leq c$ and so if  $(c',v) \not \in x^\top$ then $(c,v) \not \in x^\top$ --- i.e. if $(c,v) \in x$ then any proof of $c$ in $x$ is a proof of $c$ in $x^\top$.  
\end{proof}
By definition, if $(c,v) \in x$ then either $(c,\bullet) \in x$ or $(c,v) \in x^\top$ --- i.e. $x \leqs x^\top$.
\begin{lem}\label{iso}For any   monotone stable function $f:D(A) \rightarrow D(B)$ and state $x \in D(A)$: $\fun(\strat(f))(x) = f(x)$.
\end{lem}
\begin{proof}
Suppose $(c,a) \in \fun(\strat(f))(x)$. Then either:
\begin{itemize}
\item there exists a total $y \subseteq x$ with $((y,c),a) \in \strat(f)$, and so $(c,a) \in f(y) \subseteq f(x)$.   
\item or there exists $y \subseteq x$ with $((y,c),c') \in \strat(f)$ and $(c',\bullet) \in x$, and so $(c,\bullet) \in f(y + (c',\bullet)) \subseteq f(x)$.     
\end{itemize}
For the converse, suppose $(c,a) \in f(x)$. If $(c,a) \in f(x^\top)$ then $((x^\top,c),a) \in \strat(f)$, and so $(c,a) \in \fun(\strat(f))(x)$ as required.   

Otherwise $(c,\bullet) \in f(x)$, and by the explicit sequentiality of $f$ (Proposition \ref{KP}), there exists $c' \in A(x^\top) \cap F(y)$ such that $(c',\bullet) \in x$ and  $(c,\bullet) \in f(x^\top + (c',\bullet))$ and so $((x^\top,c),c') \in \strat(f)$ and $(c,a) \in \fun(\strat(f))(x)$ as required.     
\end{proof}
\begin{lem}For all sequential algorithms $\sigma \in D(A \Rightarrow B)$, $\strat(\fun(\sigma)) = \sigma$. 
\end{lem}
\begin{proof}
Suppose $((x,c),a) \in E(A\Rightarrow B)_\bullet$. Then:
\begin{itemize}
\item If $a \in V\!(B)_\bullet$ then $((x,c),a) \in \sigma$ if and only if  $(c,a) \in \fun(\sigma)(x)$ if and only if $((x,c),a) \in \strat(\fun(\sigma))$. 
\item If  $a  \in C(A)$ then $((x,c),a) \in \sigma$ if and only if  $(c,\bullet) \in \fun(\sigma)(x + (c',\bullet))$ if and only if      $((x,c),a) \in \strat(\fun(\sigma))$. \qedhere
\end{itemize}
\end{proof}

 To complete the proof that $\fun$ is a \emph{(bi)order-isomorphism} from $D(A \!\Rightarrow\!\! B)$ to  $[D(A),\!D(B)]$ we need to show that it is a biorder-embedding --- i.e. that 
$\fun$ and $\strat$ are  monotone in both orders (from which stability follows). Monotonicity with respect to the extensional order is straightforward, leaving the stable order.
\begin{lem}\label{stab}If $\sigma \leqs \tau$ then $\fun(\sigma) \leqs \fun(\tau)$.
\end{lem}
\begin{proof}
First, we show that  for all  $x$, $\fun(\sigma)(x) \leqs \fun(\tau)(x)$. Suppose $(c,v) \in \fun(\sigma)(x)$ but $(c,\bullet) \not\in \fun(\sigma)(x)$. Then there exists $x' \subseteq x$ with $((x',c),v) \in \sigma$ and  $((x',c),\bullet) \not\in \sigma$ and so $((x',c),v) \in \tau$ and  $(c,v)\in \fun(\tau)(x)$ as required.   
  
Now we need to show that for all $x \leqs y$, 
$\fun(\sigma)(x) =  \fun(\tau)(x) \cup \fun(\sigma)(y)$.  Suppose $(c,a) \in \fun(\sigma)(x)$, we need to show that  $(c,a) \in \fun(\tau)(x)$ or $(c,a) \in \fun(\sigma)(y)$.   By Proposition \ref{wdef}, if $(c,a) \not \in \fun(\sigma)(y)$, then $(c,\bullet) \in \fun(\sigma)(x)$, and so we may assume that $a = \bullet$. So either there  exists  an event  $((z,c),\bullet) \in \sigma$ with $z \subseteq x$  or else  there exists  $z + (c',\bullet) \subseteq x$ such that $((z,c),c') \in \sigma$. But this latter case reduces to the first one, since  either  $((z,c),c') \in \tau$ (and so $(c,\bullet) \in \fun(\tau)(x)$ and we are done), or else $((z,c),\bullet) \in \sigma$.

 Assuming $(c,\bullet) \not \in \fun(\sigma)(y)$, let $P$ be a proof of $((z,c), \bullet)$ in $\sigma$, and let $((z',c'),a)$ be the least element of $P$ such that $z' \not \subseteq y$. Then there must be an immediately preceding event in $P$ of the form   $((z'',c'),c'')$, where $z'  = z'' + (c'',V)$ for some $V$, and hence $z'' + (c'',\bullet) \subseteq x$, as $\uparrow \{x, y\}$. If $((z'',c'),c'') \in \tau$ then $(c,\bullet) \in \fun(\tau)(x)$. Otherwise $((z'',c'),\bullet) \in \sigma$ and so $(c',\bullet) \in \fun(\sigma)(y)$, and so $(c,\bullet) \in \fun(\sigma)(y)$ as required.  
\end{proof}
\begin{lem}For $f,g:D(A) \rightarrow D(B)$, if $f \leqs g$ then $\strat(f) \leqs \strat(g)$.
\end{lem}
\begin{proof}Suppose $((x,c),a) \in \strat(f)$. If $a$ is a value in $B$ then $(c,a) \in f(x) \leqs g(x)$ and so either $((x,c),a) \in \strat(g)$ or $((x,c)\bullet) \in \strat(f)$. 
If $a$ is a cell in $A$, so that $(c, \bullet) \in f(x + (a,\bullet)) = f(x) \cup g(x + (a,\bullet))$ (since $f \leqs g)$, then either $((x,c),\bullet) \in f$ or $((x,c),a) \in g$ as required.    
\end{proof}
%$\fun  for any set of states $S \subseteq D(A\Rightarrow B)$, $\fun(\bigcup S) = \bigcup_{\sigma \in S} \fun(\sigma)$ by construction --- i.e. $\fun$ is \emph{additive} --- we have shown that: 
%\begin{prop}$\fun$ is a monotone stable function from $D(A \Rightarrow B)$ to $[D(A),D(B)]$.
%\end{prop}

Hence we have shown that:
\begin{thm}\label{thm}$[D(A), D(B)] \cong D(A \Rightarrow B)$ in the category of biorders and monotone stable functions. %is an isomorphism, with inverse $\strat: D(A \Rightarrow B) \rightarrow [D(A), D(B)]$.  
\end{thm}
\begin{cor}The category of ordered concrete data structures and monotone stable functions is Cartesian closed.
\end{cor}
%Recall that an ocds is said to be stably complete if $D(A)$ is stably complete. 

\subsection{Continuous Functions and Sequential Algorithms} 
We now consider stably continuous and continuously stable functions, showing that these correspond to sequential algorithms which satisfy (respectively) the  finite-branching and  finite safety conditions. The more interesting case is the former one. We define an internal hom  $A \Rightarrow_* B$ for $\ocds$ by restricting $A \Rightarrow B$ to cells $(x,c)$ such that $x$ is a  total state in $D_*(A)$ and defining $\fun_*:D_*(A \Rightarrow_* B) \rightarrow [D_*(A),D_*(B)]$  to be the restriction of $\fun$ to states of $D_*(A \Rightarrow_* B)$ and $D_*(A)$ --- i.e.
$\fun_*(\sigma)(x) =$ $$ \{(c,a)\in E(B)_\bullet\ |\ \exists x' \subseteq x.((x',c),a) \in \sigma \vee \exists c'.(c',\bullet) \in x \wedge ((x',c),c') \in \sigma\}$$ 
Proof that this is well-defined and monotone stable follows that for $\fun$ (it is also straightforward to check that if $\sigma$ and $x$ are finite-branching then $\fun_*(\sigma)(x)$ is finite-branching.) So we need to show that for any finite-branching $\sigma$, $\fun_*(\sigma)$ is $\leqs$-continuous --- i.e. if  $\Delta \subseteq D_*(A)$ is stably directed then  $\fun_*(\sigma)(\bigvee \Delta) = \bigvee\{\fun_*(\sigma)(x)\ |\ x \in \Delta\}$. Since $\bigvee_{x \in \Delta}\fun_*(\sigma)(x) \leqs \fun_*(\sigma)(\bigvee \Delta)$ by stability of $f$, it suffices to show that for any cell $c$ in $B$, if $(c,\bullet) \in \bigvee_{x \in \Delta}\fun_*(\sigma)(x)$ then $(c,\bullet) \in \fun_*(\sigma)(\bigvee \Delta)$.
\begin{lem}\label{fb}For all $n$, either $(c,\bullet) \in \fun_*(\sigma)(\bigvee \Delta)$ or there exists a set of  events $\{((y_1,c),c_1),\ldots,((y_n,c),c_n)\} \subseteq \sigma$ such that $y_i \subseteq \bigvee \Delta$ and  $c_i = c_j$ implies $i =j$ for each $i$.
\end{lem}
\begin{proof}By induction on $n$. Suppose we have $\{((y_1,c),c_1),\ldots,((y_n,c),c_n)\} \subseteq \sigma$ satisfying the above conditions. If any of the $c_i$ is filled  with $\bullet$ in $\bigvee \Delta$, then $(c,\bullet) \in f(\bigvee \Delta)$, so suppose that for each $i \in \{1,\ldots,n\}$,  $(c_i, \bullet) \not \in  \bigvee \Delta$ and so there exists $x_i \in \Delta$ such that $(c_i,\bullet) \not \in x_i$. By directedness of $\Delta$, there exists  $x \in \Delta$ such that none of the $c_i$ are filled with $\bullet$ in $x$. Since $(c,\bullet) \in \fun_*(\sigma)(x)$, there exists an event $((y,c),a) \in \sigma$ such that $y \subseteq x$,   and either $a = \bullet$  or $a = c'$, where  $y + (c',\bullet) \subseteq x$.
If $y \subseteq \bigvee \Delta$, then in the first case  $(c,\bullet) \in \fun_*\sigma)(\bigvee \Delta)$, and in the second case we may set $y_{n+1} = y$ and $c_{n+1} = c'$, satisfying the induction hypothesis.   

If $y \not \subseteq \bigvee \Delta$, let $P$ be a  proof of $((y,c),a)$ in $\sigma$, and let $((z,d),a)$ be the least element of this proof such that $z \not \subseteq \bigvee \Delta$. Then $((z,d),a)$ is immediately preceded in $P$ by an event $((z',d),d')$ such that $z' \subseteq \bigvee \Delta$ and $z = z' + (d',V)$, for some set of values $V$. So there exists $v \in V$ such that $(d',v) \not \in \bigvee \Delta$ and hence $(d',\bullet) \in x$, since $z \subseteq x \leqs \bigvee \Delta$. 
 $((z',c),d') \geq (z',d'),d) \in \sigma$  and so we may set $y_{n+1} = z'$ and $c_{n+1} = d'$, satisfying the induction hypothesis.  
\end{proof}We can now prove that $\fun_*(\sigma)$ is  $\leqs$-continuous.
\begin{prop}\label{lvl}$\fun_*(\sigma)(\bigvee \Delta) = \bigvee_{x \in \Delta}\fun_*(\sigma)(x)$.
\end{prop}
\begin{proof}
By Lemma~\ref{fb}, if $(c,\bullet) \in \bigvee_{x \in \Delta}\fun_*(\sigma)(x)$,  either $(c,\bullet) \in \fun_*\sigma)(\bigvee \Delta)$ or there is an infinite set of events
$\{((y_i,c),c_i) \in \sigma \ |\ i \in \omega\}$ such that each $y_i \subseteq \bigvee \Delta$ and  $c_i = c_j$ implies $i =j$. In the former case we are done, so assume the latter and 
consider the tree of proofs of these events in $\sigma$. If this is infinite  branching, then either there is some $((y_i,c),c_i)$ which enables infinitely many cells $(y_j,c)$  --- in which case $c_i$ contains infinitely many values in $\bigvee \Delta$ and so $(c_i,\bullet) \in \bigvee \Delta$ and $(c,\bullet) \in f(\bigvee \Delta)$ as required --- or else there is some $y \subseteq \bigvee\Delta$ such that $y_i = y$ for infinitely many $i$ --- i.e.  $(y,c)$ contains infinitely many $c_i$ in $\sigma$, and so $((y,c),\bullet) \in \sigma$ and $(c,\bullet) \in f(\bigvee \Delta)$ as required.
 
If the tree of proofs is finite-branching then by K\"onig's lemma it contains an infinite branch --- i.e. there exists $z \subseteq \bigvee \Delta$ such that $(z,c)$ has an infinite proof in $\sigma$, and by the infinite livelock condition $((z,c),\bullet) \in \sigma$ and $(c,\bullet) \in f(\bigvee \Delta)$ as required.    
\end{proof}
 The inverse  to $\fun_*$ is the restriction of $\strat$ to stably continuous functions on finitely branching states --- i.e. for stably continuous $f:D_*(A) \rightarrow D_*(B)$, $\strat_*(f) =$
\begin{center}
$\{((x,c),a) \in C(A \Rightarrow_* B) \times V\!(B)_\bot \spc |\spc  (c,a) \in f(x)\}$\\ $\cup \{((x,c), c')  \in C(A \Rightarrow_* B) \times C(A)\spc |\spc (c,\bullet) \in f(x + (c',\bullet))\}$
\end{center}
Proof that this satisfies the safety condition and furnishes an inverse to $\fun_*$ follows the proof of Lemma~\ref{iso}. We need to show in addition that if $f:D_*(A) \rightarrow D_*(B)$ is stably continuous then $\strat_*(f)$ is finite-branching. 
\begin{lem}$\strat_*(f)$ satisfies the infinite livelock property.
\end{lem}
\begin{proof}
Suppose $((x,c),\bullet)$ has an infinite proof $ (x_\alpha, c_\alpha)_{\alpha \leq \kappa}$ in $\strat_*(f)$. If the set $\{c_\alpha\ | \ \alpha \leq \kappa\}$ is infinite then $(c,\bullet)$ has an infinite proof in $f(x)$ and so $(c, \bullet) \in f(x)$ and $((x,c),\bullet) \in \strat_*(f)$ as required. 

Otherwise, the proof contains an infinite substring $((x_{\alpha +i},c_{\alpha +i}),a_{\alpha +i})_{i < \omega}$, where $c_{\alpha+i} = c_\alpha$ and $a_{\alpha +i}$ is a cell in $A$ enabled by $x_{\alpha +i}$  for each $i<\omega$.  Let $y_i = x_{\alpha + i} + \{(a_{\alpha +j},\bullet)\ |\ j >i \wedge a_j \in A(x_{\alpha +i}) \}$ for each $i \in \omega$. The $y_i$ form a stable $\omega$-chain such that $(c_\alpha,\bullet) \in f(y_i)$ for each $i \in \omega$  and hence by stable continuity $(c,\bullet) \geq (c_\alpha,\bullet) \in \bigvee_{i \in \omega}f(y_i) = f(\bigcap_{i \in \omega}y_i) \subseteq f(x)$ and so $((x,c),\bullet) \in \strat(f)$ as required.     
\end{proof}
\begin{lem} $\strat_*(f)$ is finite-branching.
\end{lem}
\begin{proof}
%Suppose the set of $a$ such that  $\{a \in V\!(A \Rightarrow_* B) \ | \ ((x,c),a) \in \strat(f)$ is infinite. 
Suppose $(x,c)$ is an enabled cell in $A \Rightarrow_* B$ such that there are infinitely many values $a$ such that $((x,c),a) \in \strat_*(f)$. If there are infinitely many values of $B$ such that $((x,c),v) \in \strat_*(f)$ then $(c, \bullet) \in f(x)$ by the finite branching condition on $f(x)$, and so $((x,c),\bullet) \in \strat_*(f)$ as required.

Otherwise there are infinitely many cells $c'$ of $A$ such that $((x,c),c') \in \strat(f)$. Let  $Y$ be the set of states $y$ which extend $x$ by filling all but a finite number of its accessible cells with $\bullet$ --- i.e. such that  $y \leqs x$ and $A(x) - F(y)$ is finite. $Y$ is $\leqs$-directed --- if $y,y' \in Y$ then $y \cap y' \in Y$ --- and has supremum $x$.

For any $y \in Y$ there exists a cell $c' \in A(x)$ such that  $(c',\bullet) \in y$ and  $((x,c),c') \in \strat(f)$ (since there are infinitely many such cells, but only finitely many are not in $x$). By definition 
$(c,\bullet)\in f(x + (c',\bullet)) \subseteq  f(y)$. Hence $(c,\bullet) \in f(y)$ for all $y \in Y$. and thus $(c,\bullet) \in f(x)$ by continuity, and so $((x',c),\bullet) \in \strat_*(f)$ as required.  
\end{proof}

%\subsection{Concrete Data Structures and Continuously Stable Functions} 
Finally, we observe that $A \Rightarrow_* B$ is also the internal hom in the category of ocds and continuously stable functions,  based on the following observations:
\begin{lem}\label{alg}For any ocds $A$, $D^*(A)$ is algebraic.
\end{lem}
\begin{proof}Let $\K(A)$ be the set of finitely generated, finitely safe states of $A$ --- i.e.  $x \in \K(A)$ if and only if it is the upper set of the union of a finite set of proofs. It is straightforward to show that every such state is $\subseteq$-compact and that every $y \in D^*(A)$ is the union of $\K(y)=  \{x \in \K(A)\ |\ x \subseteq y\}$.  
\end{proof}
Thus $f:D^*(A) \rightarrow D^*(B)$ is continuous if and only if for any $y \in D^*(A)$ and $e \in f(y)$ there exists $x \in \K(y)$ such that $(c,a) \in f(x)$, and we may define an internal hom $(A \Rightarrow^* B)$ by restriction of $A \Rightarrow B$ to cells $(x,c)$ such that $x$ is finitely generated, and  
$\fun^*:D^*(A \Rightarrow_* B) \rightarrow [D^*(A),D^*(B)]_{\EB}$ by restriction of $\fun$ to finitely safe states --- i.e. 
$\fun^*(\sigma)(x) =$ $$ \{(c,a)\in E(B)_\bullet\ |\ \exists x' \in \K(x).((x',c),a) \in \sigma \vee \exists c'.(c',\bullet) \in x \wedge ((x',c),c') \in \sigma\}$$ 
% \{(c,a)\in E(B)_\bullet\ |\ \exists x' \in \K(x).((x',c),a) \in \sigma\}$, 
for which stability follows as for $\fun$, and continuity follows from proposition \ref{alg}. Its inverse,  $\strat^*:  [D^*(A),D^*(B)]_{\EB} \rightarrow D^*(A \Rightarrow^* B) $ is similarly defined by restriction of $\strat$ to events $((x,c),a)$ such that $x$ is finitely generated, which will have a finite proof in $\strat^*(f)$ provided $(c,a)$ has a finite proof in $f(x)$. The composition of $\strat^*$ with $\fun^*$ in each direction is the identity on all compact elements, and is therefore the identity on all elements. Finally, we may show (by induction on proof-length) that:
\begin{lem}\label{pf}$((x,c),a) \in E(A \Rightarrow_* B)$ has a finite proof if and only if $x \in \K(A)$ and $c$ has a finite proof in  $B$. 
\end{lem}
Hence the ocds $A \Rightarrow^* B$ and $A \Rightarrow_* B$ are isomorphic in $\ocde$.

%\section{Denotational Semantics of \espcf}
\section{Full Abstraction and Universality}
\label{sec:s7}
We now return to the may and must testing semantics of bounded and unbounded \espcf.  In each case we now have an intensional representation of each program  of pointed type  $t:P$ as a state  $\(t\)$ of an ordered concrete data structure $\(P\)$, since the biorder denoted by a type of the form $\nt \rightarrow T$
 is isomorphic to $\[T\]^\omega$.  (Formally, \espcf may be interpreted by applying the  \emph{families construction} (biproduct completion)  to our categories of ordered concrete data struction --- see  \cite{AMV}.) 
%We  describe explicitly the sequential algorithm denotations of some terms in the  (must testing) model of unbounded nondeterminism in Section~\ref{sec:s6}.2, before showing that this model is not fully abstract by identifying a weak continuity property of definable terms, relating the stable and extensional orders.

First, our aim is to   show that our models of bounded \espcf  are fully abstract  with respect to may testing and must testing, respectively, and characterize their  \espcf-definable elements (a form of \emph{universality} result). 
  Full abstraction for models in which terms denote continuous morphisms may be established by proving that each type-object has a basis  of elements which are definable (i.e. denotations of terms).  In the case of bounded \espcf one  may derive such a basis from the fact --- already remarked (Proposition \ref{univunary}) --- that all elements of finite types are definable in $\Lambda^+$, since every pointed  \espcf type $T$ is a limit of a chain of finite products of such types (see \cite{csl09} or \cite{McCT} for details of such a proof). However, this rather cheap reduction to universality  for $\Lambda^+$ is not very informative about the definability properties of infinite elements --- such as the difference between the may and must testing models. We shall prove  a stronger result extending Proposition \ref{univunary} --- that the type $U \triangleq  \nt \rightarrow \o \rightarrow \o$ is \emph{universal} in  our models of  bounded \espcf --- i.e. $T \unlhd U$ for all pointed \espcf types $T$. In particular $U \rightarrow U \unlhd U$ --- i.e. $U$ denotes a  \emph{reflexive object} in the categories of concrete data structures/complete biorders and continuous functions.
\begin{rem} This is similar to results for the sequential algorithms model of deterministic SPCF \cite{bist}, which are refined in \cite{afflong} to show that all retractions are definable in a linearly typed version of SPCF without nesting of function calls (and fixed point combinators replaced with iteration). The retractions given here can be defined  within a similar typing system for non-deterministic SPCF %  --- to   contain and simplify exposition, we  shall not give details here  
 \end{rem}
\begin{rem}We have already observed that $\[U\] \cong D^*(\U)$, where $\U$ is the ocds with a  countable set of initial cells $\{c.i\ |\ i \in \omega\}$ which may each be filled by a single value, and that this is a universal object in  the categories of countable  ocds and stably continuous/continuously stable functions. Thus every pointed type  of  \espcf  denotes a retract  of $\U$. It remains to show that each such a retraction is definable. 
\end{rem}

The proofs that there are \espcf-definable retractions from pointed \espcf types into $U$ are similar for the may testing and must testing models --- we give the former (slightly simpler) case first.

Observe that $N \rightarrow T$ denotes a $\Na$-indexed product (or infinite list) of copies of $\[T\]$. Thus we may define the following list  operations on terms:\\
 Given $t_1:T$ and $t_2:\nt \rightarrow T$, let $t_1 \!\!::\!\!t_2:\nt \rightarrow T \triangleq \lambda u.(\eq(u,0)\spc t_1)\spc (\pred(u)\spc t_2)$ and given $t: \nt \rightarrow T$, define $\hd(t):T \triangleq t\spc {\mathtt{0}}$ and $\tl(t) \triangleq \lambda v.t\spc \suc(v)$, so that $\[\hd(t_1\!\!::\!t_2)\] = t_1$ and $\[\tl(t_1\!\!::\!t_2)\] = \[t_2\]$. 
Thus, in particular, since $\nat \triangleq (\nt \rightarrow \o) \rightarrow \o$
\begin{lem}\label{lc}$o \rightarrow \nat \unlhd \nat$. 
\end{lem} 
For any type $T$, define the pairing and projections:\\
 $\pair: (\nt \rightarrow T) \rightarrow (\nt \rightarrow T) \rightarrow \nt \rightarrow T \triangleq \Y\lambda F.\lambda x.\lambda y.\hd(x)\!\!::\!\!(F\spc y \spc \tl(x))$, with
  $\fst:(\nt \rightarrow \o) \rightarrow (\nt \rightarrow \o) \triangleq \Y\lambda F.\lambda x.(\hd(x)\!\!::\!(F\spc \tl(\tl(x)))$ and $\snd:(\nt \rightarrow \o) \rightarrow (\nt \rightarrow \o) \triangleq \Y\lambda F.\lambda x.\hd(\tl(x))\!\!::\!\!(F\spc \tl(\tl(x))$, so that $\[\fst\spc \pair (t_1,t_2)\] = \[t_1\]$ and    
 $\[\snd\spc \pair (t_1,t_2)\] = \[t_2\]$.
\begin{lem}\label{ld}For any type $T$,  $\nt \rightarrow \nt \rightarrow T \unlhd \nt \rightarrow T$.
\end{lem}
\begin{proof}
  Let:

  \smallskip
\noindent {\small
  $\merge:(\nt \rightarrow \nt \rightarrow T) \rightarrow (\nt \rightarrow T)  \triangleq \lambda f.(\Y\lambda F.\lambda u.\pair((f\spc u),(F\spc f)\spc \suc(u)))\spc {\mathtt{0}},$\\
  $\unmerge:(\nt \rightarrow T) \rightarrow \nt \rightarrow \nt \rightarrow T \triangleq (\Y\lambda F.\lambda u.\lambda x.\lambda v. (\eq(u,v)\spc \fst(x))\spc ((F\spc \suc(u))\spc \snd(x)))\spc {\mathtt{0}}.$
  }
\end{proof}
%We assume \espcf-definable pairing and projection operations on expressions of type $\nt$, such that  $\[\fst(\pair(s,t))\] = \[s\]$  and $\[\snd(\pair(s,t))\] = \[t\]$. Thus for all pointed types $T$, 
%$\nt \rightarrow \nt \rightarrow T \unlhd \nt \rightarrow T$  via the retraction $(\lambda f.\lambda x.(f \fst(x))\spc \snd(x),\lambda g.\lambda y.\lambda z.g\spc \pair(y,z))$.  
So in particular,  $\nt \rightarrow U \unlhd U$. Recall that the denotation of the type $\nat$ is isomorphic to the biorder $\xp(\Na)_\up$.
\begin{lem}\label{natretrac}$\nat \unlhd U$.
\end{lem}
\begin{proof}Observe that we may faithfully represent any subset $X \subseteq \Na$ as a function $b \in \[U\]$ such that $b(i) = I$ if $i \in X$  and $b(i) = \ebot$ otherwise (where $I \in [\Sigma,\Sigma]$ is the identity function).
 %This corresponds to a retraction $\ij: \[\nat\] \unlhd  \U:\pj$, where:
%\begin{itemize}
%\item $\ij(a)(i) = \bots$ if $a = \bots$, $\ij(a)(i) = I$ if $a(i) =  and $(c.i,*) \in \ij(a)$ iff $(c,i) \in a$
%\item $\pj(b) = \bigsqcup_{i \in \Na}b( $ iff  there exists $i$ such that $(c.i,\bullet) \in b$,\\ 
%$(c,i) \in \pj(b)$ iff $(c.i,*) \in b$ or there exists $j$ such that $(c.j,\bullet) \in b$,\
This corresponds to the following \espcf-definable retraction: 
$$\inj:\nat \rightarrow U  \triangleq \lambda f.\lambda u.\lambda x.f\spc \lambda v. (\eq(u,v)\spc x)\spc \Omega$$ 
$$\pj:U \rightarrow \nat \triangleq  \lambda f.\lambda g.\rnd \spc \lambda w.(f\spc w) \spc (g \spc w)$$
For any $g: \widetilde{\Na} \rightarrow \Sigma$,  the set $\{\tops[g(n)]_n \ |\ n \in \Na\}$ is stably bounded above by the constantly  $\tops$ function, and $g = \bigwedge \{\tops[g(n)]_n \ |\ n \in \Na\}$. Thus $(\pj\spc \inj (f))(g) = $  
$\bigwedge_{n \in \Na}f(\tops[g(n)]_n) = f(\bigwedge_{n \in \Na}\tops[g(n)]_n) = f(g)$ by stability.  
\end{proof}

The key retraction reducing the \emph{order} of types (depth to which the arrow is nested on the left) is based on the observation  that we may represent a continuously stable   function $f$  from $\[U\] \cong [\Sigma, \Sigma]^\omega$ into $\Sigma$ via the following information:
\begin{itemize}
\item Its set of sequentiality indices,
\item For each sequentiality index $i$, the function $\lambda x.f(x[I]_i)$.
\end{itemize}
\begin{lem}\label{s}For all $g \in \[\nt \rightarrow \o \rightarrow \o\]$, $f(g) = \bigsqcup_{i \in \si(f)}(g(i)\spc f(x[I]_i))$.
\end{lem}
\begin{proof}
If $g(i) = \sbot = \etop$ for some sequentiality index $i \in \si(f)$ then $f(g) =  \sbot =  g(i)\spc f(g[I]_i) = \bigsqcup_{i \in \si(f)}(\pi_i(x)\spc f(x[I]_i))$. So suppose that $g(i) \in \{\ebot,I\}$ for all $i \in \si(f)$.\\
If $g(i) = \ebot$ then  $g(i) \spc f(g[I]_i) = \ebot$ and if  $g(i) = I$  then  $g(i) \spc f(g[I]_i) = I\spc f(g) = f(g)$.\\
 So if $g(i) = \ebot$ for every sequentiality index $i$, then $ \bigsqcup_{i \in \si(f)}(g(i)\spc f(x[I]_i)) = \ebot = f(g)$ by Lemma~\ref{si}.\\
Otherwise there is some $i \in \si(f)$ such that  $g(i) = I$  and so  $\bigsqcup_{i \in \si(f)}g(i)\spc f(g[I]_i) = f(g)$ as required.
% $g(i) \spc f(g[I]_i) = I\spc f(g) = f(g)$ and soas required.
\end{proof}

Thus, extracting sequentiality indices for $f,\{\lambda x.f(x[I]_i) \ |\ i \in \si(f)\}, \{\lambda x.f(x[I]_i[I]_j)\ |\ i \in \si(f),j \in \si(\lambda x.f(x[I]_i)\}, \ldots$ generates a set of finite sequences of  indices from which $f$ may be recovered.   
\begin{lem} \label{retrac}$U \rightarrow \o \unlhd \nt \rightarrow \nat$.
\end{lem}
\begin{proof}Let:
\begin{center}
$\inj: (U \rightarrow \o) \rightarrow  \nt \rightarrow \nat \triangleq  \lambda f.\Y_{\nt \rightarrow \nat}\lambda F.(\lambda x.f\spc \lambda y.x)\!\!::\!\!(\merge(\lambda u.F\spc \lambda x.f\spc x[I]_{u}))$.
 $\proj:(\nt \rightarrow \nat) \rightarrow U \rightarrow \o \triangleq \Y_{U}\lambda F.\lambda x.\lambda y.\hd(x)\spc \lambda v.(y\spc v)\spc  (F\spc (\unmerge(\tl(x))\spc v))$.   
\end{center}
We prove  by induction on $k \in \Na$ that  for all $e \in \[\nt \rightarrow \o \rightarrow \o\]$, if $|\{ n \in \nat \ |\  e(n) = I\}| \leq  k$  then for all $f \in \[U \rightarrow \o\]$,  $\inj(\proj(f))(e) =  f(e)$. For the induction step, unfold the fixed point, so that:
$\inj(\proj(f))(e) = \bigsqcup_{i \in \si(f)}e(i)\spc \inj(\proj(\lambda x.f(x[I]_i)))(e)$. Thus, by Lemma~\ref{s} it is sufficient to show that if $i \in \si(f)$ then $e(i)\spc \inj(\proj(\lambda x.f(x[I]_i)))(e) = e(i)\spc f(e[I]_i)$. This is evident if $e(i) \not = I$ (i.e. $e(i)$ is  the constant function $\sbot$ or $\ebot$). If $e(i) = I$ then \\  $f(e[I]_i) = \inj(\proj(\lambda x.f(x[I]_i))(e[\ebot]_i)\\  \sqleq \inj(\proj(\lambda x.f(x[I]_i))(e) \\\sqleq \inj(\proj(\lambda x.f(x[I]_i))(e[\etop]_i) = f(e[I]_i)$ by induction hypothesis applied to $e[\ebot]_i$ and $e[\etop]_i$. \\
So  $e(i)\spc \inj(\proj(\lambda x.f(x[I]_i))(e) = e(i)\spc f(e[I]_i)$,  completing the induction. 

Since any element $e \in \[U\]$ is the $\sqleq$-supremum of the approximants $e_k$ such that $e_k(i) = e(i)$ if $i \leq k$ and $e_k(i) = \ebot$  otherwise  (to which the above induction applies), for any $f \in \[U \rightarrow \o\]$,  $\inj(\proj(f))(e) = f(e)$ by continuity of $f$. 
\end{proof}
By composing definable retractions we may now show that $U$ is a (definably) reflexive object (i.e. $U \rightarrow U$ is a definable retract of $U$). 
\begin{prop}\label{refl}$U \rightarrow U \unlhd U$.
\end{prop}
\begin{proof}We have $U \rightarrow U = U \rightarrow \nt \rightarrow o \rightarrow o \cong \nt \rightarrow \o \rightarrow  U \rightarrow \o$ \\
$\unlhd \nt \rightarrow \o \rightarrow \nt \rightarrow  \nat$\ \ \ by Lemma~\ref{retrac} \\
$ \cong \nt \rightarrow  \nt \rightarrow \o \rightarrow \nat \unlhd \nt \rightarrow \nt \rightarrow \nat$ by Lemma~\ref{lc}\\   
$\unlhd \nt \rightarrow \nt \rightarrow U$ by Lemma~\ref{natretrac}\\
 $\unlhd \nt \rightarrow U  \unlhd U$ by Lemma~\ref{ld}.
\end{proof}
\subsection{Full Abstraction}
Remarking that $\U$ also  corresponds to a universal object in the category of countable ocds and continuous functions, we may apply largely the same argument to establish that $U$ is also a reflexive type in our must testing model of bounded \espcf: most of the terms used in the proof of Proposition \ref{U} also denote retractions in $\SB$, and the same proofs apply. The exception is the retraction of $\nat$ into $U$  (Lemma~\ref{natretrac}), which  requires the use of unbounded nondeterminism. Recall that in our stably continuous model, $\nat$ corresponds to a powerdomain of \emph{finite} sets of natural numbers: by using a different representation of this biorder in $\U$, we may  construct an alternative retraction.    
\begin{lem}$\nat \unlhd U$ in the must testing model.
\end{lem}
\begin{proof}As in Lemma~\ref{natretrac} we may  represent a (finite) set $X$ of integers as an element $b \in \[U\]$  such that $\pi_i(b) = I$ if $i \in X$ and $\pi_i(b) = \etop$ otherwise. The problem is that the projection which sends $b \in \[U\]$ to the set of all natural numbers  $i$ such that $b(i) = I$ exhibits unbounded non-determinism (since there may be infinitely many such $i$)  and is therefore not stably continuous. Our solution is to use a second value $c \in \[U\]$ to represent an upper bound for~$X$~--- i.e. $c(i) = I$ if there exists $c \in X$ such that $i \leq j$, and $c(i) = \etop$ otherwise.    We can then return all natural numbers $i$ such that $b(i) = I$ and $\forall j \leq i.c(j) = I$  by stepping through values of $i$ in sequence.

Thus  we have a definable retraction of $\nat$ into $U^2$, given by the terms:
\begin{center}
$\inj_1:\nat \rightarrow U \triangleq \lambda x:\nat.\lambda u:\nt.\lambda y:\o.x\spc \lambda v.(\eq(u,v)\spc y)\spc \e$  
$\inj_2:\nat \rightarrow U \triangleq \lambda x:\nat.\lambda u:\nt. \lambda y. {\mathtt{leq}}(u,v)\spc y)\spc \e$.     
$\proj:U \rightarrow U \rightarrow \nat \triangleq \lambda y.\lambda z. \lambda k. (\Y_{\nt \rightarrow \o}\lambda F.\lambda v:\nt.(z\spc v) \spc (((y\spc v)\spc (k\spc v)) \orr (F\spc \suc(v))))\spc 0$ 
\end{center}
  \noindent Where:

  \smallskip
\noindent ${\mathtt{leq}}\!:\!\nt \rightarrow \nt \rightarrow \boo \triangleq \lambda u.\lambda v.\lambda x.\lambda y.(\Y\lambda f.\lambda w.(\eq(u,w)\spc x) ((\eq(w,v)\spc y) \spc (f\spc \suc(w))))\spc {\mathtt{0}}$.   
%(We assume primitive recursive operations on $\nt$ allowing us to define ${\mathtt{leq}}(n,m):\boo$, which  evaluates to $\tru \triangleq \lambda x.\lambda y.x$ if $n \leq m$ and   $\ff \triangleq \lambda x.\lambda y.x$ otherwise.) 
\end{proof}
Thus $U \unlhd U \rightarrow U$ in the must testing model, exactly as in Proposition \ref{refl}, and we may prove by  structural induction:
\begin{prop} \label{U}For every pointed type $P$, $P \unlhd U$ in the may testing and must testing semantics of bounded \espcf.
\end{prop} 
It remains to prove full abstraction for both models. Say that completeness holds at type $T$ if for all closed $s,t:T$, if $s \lmay t$ then $\[s\]_{may} \sqleq \[t\]_{may}$ and if  $s \lmust t$ then $\[t\]_{must} \sqleq \[s\]_{must}$.
\begin{lem}Completeness holds at the type $U \triangleq \nt \rightarrow \o \rightarrow \o$. 
\end{lem}
\begin{proof}Suppose e.g. $s \lmust t$. Then by soundness and adequacy, for any $d \in \Na$ and $e \in \{\tops,\sbot\}$  we have $(\[s\]_{must} \spc d)\spc e = \tops = \ebot$ implies  $(\[t\]_{must}\spc d)\spc e = \tops = \ebot$, and so  $\[t\]_{must} \sqleq \[s\]_{must}$.
\end{proof}
Thus completeness holds at all pointed types (it also holds at  $\nt$, since e.g. if $t \lmay s$ then $(((\eq\spc t)\spc {\mathtt n})\spc \e)\spc \Omega \Downarrow^\may$ implies $(((\eq\spc s)\spc {\mathtt n})\spc \e)\spc \Omega\Downarrow^\may$, and hence  $\[t\]_{may} = \[s\]_{may}$).
%So we have extended inequational completeness to all types and proved full abstraction.
\begin{thm}For all terms $s,t$, $s\lmay t$ if and only if $\[s\]_{may} \sqleq  \[t\]_{may}$ and $s \lmust t$ if and only if  $\[s\]_{must} \sqsubseteq \[t\]_{must}$.  
\end{thm}
\begin{rem}
Universality of $U$ has several other applications. For example, it allows us to characterize the \espcf-definable infinitary elements of a (pointed) type-object $\[T\]$ as those elements $e \in \[T\]$ such that $\inj(e):U$ corresponds to a computable (partial) function from $\Na$ into $\{I,\top\}$.  As $\U$ is a reflexive object, endomorphisms on $\U$ form a $C$-monoid, providing an interpretation of the untyped $\lambda$-calculus, similar to other graph models, but with a weakly sequential, intensional interpretation as sequential algorithms.   Since $[\U, \U_\up] \unlhd \U$ and $[\U,\U]_\up\unlhd  \U$, the $C$-monoid of endomorphisms on $\U$ also furnishes weakly sequential semantics for the untyped call-by-value  and lazy $\lambda$-calculi, to which we can apply the argument in \cite{ufpc} to prove full abstraction.
\end{rem}

\subsection{Unbounded Nondeterminism}
The situation in the non-continuous models  is different. Although finitary elements of these models are still definable, in the absence of continuity this does not imply full abstraction, which in fact does not hold (contradicting  \cite{fos06}). We focus on the model of must testing (since this requires non-continuity).
We have shown by example that some events with infinite proofs are required to distinguish inequivalent terms in this model, but hinted that there are many more  which may distinguish denotations of terms which cannot be separated operationally --- we now give an example of a  stable and monotone function  which can be used to distinguish two must equivalent terms, establishing that the model is not fully abstract.
 Consider the function  $h:\[\nat\] \rightarrow \[\o \rightarrow \o\]$ such that:
\begin{itemize}
\item $h(x) = \etop$ if $x = \etop = \sbot$ ($\si(x)$ is empty),
\item $h(x) = \ebot$, if $\si(x)$ is finite but non-empty,
\item $h(x) = I$, otherwise ($\si(x)$ is infinite). 
\end{itemize}
This is monotone and stable, since $\si(x) \subseteq \si(y)$ and $\si(y)$ finite implies $\si(x)$ finite, and    $x \leqs y$ in $\[\nat\]$ implies $x =y $ or $x=\sbot = \etop$. However, it cannot be computed in \espcf, since there is no test which can check for unboundedness and then return, rather than diverging. Note that  $g:\[\nat\] \rightarrow \[\o\]$ such that $g(x) = \ebot$ if $\si(x)$ is finite, and $g(x) = \etop$ otherwise, is definable as $\lambda x.(\Y_{\nt \rightarrow \o} \lambda f.\lambda u.x\spc \lambda v.({\mathtt{leq}}(u,v)\spc \e)\spc (f\spc \suc(v)))\spc {\mathtt{0}}$.

We may prove that $h$ is not definable in \espcf by showing that all definable functions have the following property:
\begin{defi}A monotone stable function of extensionally complete biorders $f:D \rightarrow E$ is \emph{weakly continuous} if for any $\sqleq$-directed set $X$, $f(\bigsqcup X) \leqs \bigsqcup f(X)$.
\end{defi}
The function  $h$ is not weakly continuous: let $X$ be the  $\sqleq$-directed set $\{\bigsqcup_{i \in J}\pi_i:\Sigma^\omega \rightarrow \Sigma\ |\ J \subseteq_{\mathit{fin}} \Na\}$. Then  $\si(\bigsqcup X) = \Na$ and so $h(\bigsqcap X) = I \not \leqs \ebot   = \bigsqcup_{x \in X} h(x)$. However,  all \espcf terms denote weakly continuous functions.  We prove this using a logical relation  --- for each type $T$, we define a set  $H(T) \subseteq |\[T\]|$ of \emph{hereditarily weakly  continuous} elements of $\[T\]$ as follows: 
\begin{itemize}
\item $H(\nt) = \Na$ and $H(\o) = \Sigma$
\item $f \in H(S \rightarrow T)$  if $f$ is weakly continuous and for any $x \in H(S)$, $f(x) \in H(T)$.    
\end{itemize}
A function $f:\[S_1\] \times \ldots \times\[S_n\] \rightarrow \[T\]$  is hereditarily weakly continuous  if it is weakly continuous and for any $\langle x_1,\ldots,x_n \rangle \in H(S_1) \times \ldots \times H(S_n)$, $f(x_1,\ldots,x_n) \in H(T)$.
\begin{prop}\label{wc2}Every term of \espcf denotes a hereditarily weakly continuous function.  
\end{prop}
\begin{proof}
A straightforward structural induction. For the fixed point operator, we note that if  $F$ is a  $\leqs$-directed set of weakly continuous functions from $D$ to (bicomplete) $E$, then $\bigvee F$ is weakly continuous, since for any $\sqleq$-directed set $X$, 
$(\bigvee F)(\bigsqcup X) = \bigvee\{f(\bigsqcup X)\ |\ f \in F\} \leqs \bigvee\{\bigsqcup_{x \in X} f(x)\ |\ f \in F\}$. 
\end{proof}
Hence we have established that:
\begin{prop}There is no term $t:\nat \rightarrow \o \rightarrow \o$ which denotes the  function $h$. 
\end{prop}
Moreover,  the weak continuity of definable functions may be used to prove that the model is not fully abstract: we now give  a \espcf-definable
 test which distinguishes $h$ from all weakly continuous functions, and therefore from all definable elements of the model.
Define $\bchoice: \nt \rightarrow \nat \triangleq  \lambda u. \lambda f.?_N\spc \lambda v.({\mathtt{leq}}(u,v)\spc (f\spc v))\spc \e$  so that $\bchoice(n)$ returns a bounded choice over all values less than $n$. Clearly, $?(n)_{n \in \omega}$ is a $\sqleq$-chain with $\bigsqcup_{n \in \omega}?(n) = ?_N$.  

\begin{lem}For any weakly continuous $f:\[\nat\] \rightarrow \[\o \rightarrow \o\]$, $f(?_N)\bigsqcup_{i \in \omega}(f(?(i))(\bot_S)) = f(?_N)(\sbot)$. 
\end{lem} 
\begin{proof} If $f(?_N)$ is a constant function, this is evident. Otherwise $f(?_N) = I$. By weak continuity $I = f(?_N) \leqs \bigsqcup_{n \in \omega}f(?(n))$, and by extensional monotonicity $ \bigsqcup_{i \in \omega}f(?(i)) \sqleq f(?_N) = I$, and so $\bigsqcup_{n \in \omega}f(?(n)) = I$. So $f(?_N)\bigsqcup_{n \in \omega}(f(?(n))(\sbot)) =  I(I(\sbot) =  \sbot =  f(?_N)(\sbot)$ as required.
\end{proof}
We now give \espcf terms of type $(\nat \rightarrow \o \rightarrow \o) \rightarrow \o$  corresponding to this test. Let
$t \triangleq \lambda f.(f\spc ?_N)\spc (?_N\spc \lambda  u.(f\spc  \bchoice(u))\spc \Omega)  $ and  $s = \lambda f.(f\spc ?_N)\spc \Omega$. 
\begin{lem}$\[t\]_\must \not = \[s\]_\must$. 
\end{lem}
\begin{proof}
Observe that $h(?_N) = I$ but $h(?(n))$ is the constantly $\ebot$ function for each $n$, and hence $\[t\](h) = h(?_N)(\bigsqcup_{n \in \omega} h(?(n)(\sbot))) = \ebot$ but $\[s\](h) = h(?_N)(\sbot)  = \sbot$.
\end{proof}
\begin{lem}$t$ and $s$ are  must equivalent in \espcf.
\end{lem}
\begin{proof}
By extensionality of \espcf, it suffices to show that if $r:\nat \rightarrow \o \rightarrow \o$ is any closed term of \espcf then  $t\spc r$ is equivalent to $s\spc r$. By Proposition \ref{wc2}
$r$ denotes a  weakly continuous function 
and so $\[t\spc r\] = \[r\](?_N)(\bigsqcup_{i \in \omega}\[r\](?(i))(\sbot)) = \[r\](?_N)(\sbot) = \[s\spc r\]$, and so by computational adequacy, $t\spc r \Downarrow_\must\mho$ if and only if $s\spc r \Downarrow_\must \mho$ as required.
\end{proof}
Thus we have shown that:
\begin{thm} The  must testing  semantics of \espcf in $\B$ is not fully abstract. 
\end{thm}
%Returning to example \ref{exa} --- of terms of type $(U \rightarrow \o) \rightarrow o$ which require  

%$p$ and $r$  are distinguished by application to $t$ ---
% $p \spc t \not \Downarrow_\must$ and $r\spc t \Downarrow_\must$, since $\(p\)$ contains the event  $((Q_\omega,c),P_\omega)$ --- which has the proof $((Q_k,c),P_k)_{k \leq \omega}$ in $\(p\)$ --- but  $r$ does not, although every event in $\(p\)$ with a finite proof also occurs in $\(r\)$.  

\subsection{Further Directions}

We have established new extensional and intensional representations of higher-order functional computation with bounded and unbounded nondeterminism, and shown how they can be used to interpret a simple but expressive programming language. Further study of this model may shed light on the ongoing debate over the relevance of the concept of \emph{fairness} \cite{djikstra}, or bear relevance to abstract interpretation of functional languages.  

In this paper, we have considered only simply typed languages, but a more general domain theoretic approach would extend our results to recursive types. As well as fixed points for morphisms, we may also construct fixed points (minimal invariants) for functors on our categories of (complete)  biorders, as coinciding limit/colimits of ordinal chains of embedding projection pairs (in the continuous cases, these will be $\omega$-chains) obtained by standard constructions. We  refer to \cite{ufpc} for one particular application --- the construction of fully abstract models of call-by-value and lazy call-by-name $\lambda$-calculi by solving the domain equations  $D \cong [D,D_\up]$ and $D \cong [D,D]_\up$ (respectively) in the category of stably complete biorders and stably continuous functions (the models described  \cite{ufpc} are in the category of bicpos \cite{Be}, but all go through in $\SB$).   
\begin{thm}[\cite{ufpc}]The semantics of $\lambda_v$ in  $\SB$ is fully abstract.
\end{thm}
By our results relating stably continuous functions to sequential algorithms we may therefore construct fully abstract intensional  models of $\lambda_v$ and $\lambda_l$ which (unlike e.g. the games model of the lazy $\lambda$-calculus in \cite{AMl}) is fully abstract without any ``intrinsic preorder quotient''. We leave as further work the extension of our semantics of unbounded non-determinism to recursive types. 
% However,  they leave  something to be desired --- due to the construction of the limit, even compact elements are presented as infinite sequences.  Our 

Further questions suggested by our semantics include:
\begin{itemize}
\item Can the notion of weak continuity be completed to give a characterization of the functions and/or sequential algorithms which are definable in unbounded \espcf ? We have also left open some important questions regarding the algebraicity properties of biorders arising from ocds, which might suggest further constraints on them.
\item Can we interpret nondeterministic languages with richer type systems (for example, explictly controlling the use of effects such as nondeterminism). In particular, can we relate powerdomain constructions on biorders to ordered concrete data structures.
\item Can we describe models of unbounded non-determinism which make the ``games-like'' structure of ordered concrete data structures more explicit --- for example, using graph games \cite{HyS} or concurrent games \cite{RW} ? Is there a corresponding linear decomposition of our model?   
\item Jones et al. \cite{ninjon} have described powerful (but necessarily incomplete) methods for deciding properties of terms in weakly sequential languages such as the untyped call-by-value $\lambda$-calculus --- can we relate these to our intensional and extensional models? 
\end{itemize}
%\bibliography{names}

\end{document}